\documentclass[12pt]{article}
\usepackage[utf8]{inputenc}
\usepackage{amsfonts,amsmath,amssymb,mathtools}
\usepackage[left=2cm,right=2cm,top=2cm,bottom=2cm,bindingoffset=0cm]{geometry}
\usepackage{graphicx}
\usepackage{cite}
\usepackage{authblk}
\usepackage{xcolor}
\usepackage{hyperref}
\usepackage{cancel}
\usepackage{subcaption}

\newcommand\bem{\begin{pmatrix}}
\newcommand\eem{\end{pmatrix}}
\newcommand\beq{\begin{equation}}
\newcommand\eeq{\end{equation}}
\newcommand\beqs{\begin{equation*}}
\newcommand\eeqs{\end{equation*}}
\newcommand{\tr}{\text{tr}}
\newcommand{\pd}{\partial}

\newcommand{\mO}{\mathcal{O}}

\newcommand{\mT}{\mathcal{T}}
\newcommand{\mD}{\mathcal{D}}

\DeclareMathOperator{\csch}{csch}

\numberwithin{equation}{section}

\hypersetup{
    colorlinks=true,
    linkcolor=black,
    citecolor=blue,   
    urlcolor=blue,
}

\title{\bf Classical and quantum butterfly effect in nonlinear vector mechanics}
\author[1,2,3]{Nikita~Kolganov\thanks{\href{mailto:nikita.kolganov@phystech.edu}{nikita.kolganov@phystech.edu}}}
\author[1,2]{Dmitrii~A.~Trunin\thanks{\href{mailto:dmitriy.trunin@phystech.edu}{dmitriy.trunin@phystech.edu}}}
\affil[1]{Moscow Institute of Physics and Technology, 141700, Institutskiy pereulok, 9, Dolgoprudny, Russia}
\affil[2]{Institute for Theoretical and Experimental Physics, 117218, B. Cheremushkinskaya, 25, Moscow, Russia}
\affil[3]{Institute for Theoretical and Mathematical Physics, Moscow State University, 119991, Leninskie Gory, GSP-1, Moscow, Russia}
\date{\today}

\begin{document}

\maketitle

\begin{abstract}
We establish the correspondence between the classical and quantum butterfly effects in nonlinear vector mechanics with the broken $O(N)$ symmetry. On one hand, we analytically calculate the out-of-time ordered correlation functions and the quantum Lyapunov exponent using the augmented Schwinger-Keldysh technique in the large-$N$ limit. On the other hand, we numerically estimate the classical Lyapunov exponent in the high-temperature limit, where the classical chaotic behavior emerges. In both cases, Lyapunov exponents approximately coincide and scale as $\kappa \approx 1.3 \sqrt[4]{\lambda T}/N$ with temperature $T$, number of degrees of freedom $N$, and coupling constant $\lambda$.
\end{abstract}

\newpage


\section{Introduction}
\label{sec:intro}

In a classical chaotic system, a small perturbation in initial conditions leads to an exponential divergence of trajectories, $\| \delta \mathbf{z}(t) \| \sim e^{\kappa_{cl} t} \| \delta \mathbf{z}(0) \|$. Here, $\delta \mathbf{z}$ denotes the distance between two trajectories in a phase space with respect to a norm $\| \cdot \|$, and the positive real number $\kappa_{cl}$ is called the maximal Lyapunov exponent. Such a sensitivity to initial conditions is broadly known as the butterfly effect. This effect occurs in numerous classical dynamical systems, lays the foundation for the classical thermodynamics and hydrodynamics, and has been extensively studied since its discovery in 1963~\cite{Lorenz-1, Lorenz-2, Chernov-1, Ott}.

On the contrary, quantum chaos and quantum butterfly effect are more subtle and less studied than their classical counterparts because the uncertainty principle prohibits infinitesimal shifts of trajectories and makes it impossible to define the quantum Lyapunov exponent directly. Instead, one introduces alternative diagnostics that are well-defined in quantum case and distinguish integrable and chaotic systems in the semiclassical limit. The oldest and most famous example of such a diagnostic is the statistics of energy level spacings~\cite{Bohigas, Berry, Haake, Ott, Stockmann}. There are also definitions of quantum chaos that rely on the calculation of dynamical entropy~\cite{Connes, Alicki}, decoherence~\cite{Zurek}, entanglement~\cite{Valdez, Nie}, out-of-time ordered correlation functions~\cite{Larkin, Maldacena-bound, Kitaev-talks, Swingle-popular}, spectral form factor~\cite{Cotler, Koch, Ma, Khramtsov}, Krylov complexity~\cite{Parker, Avdoshkin, Gorsky, Smolkin, Rabinovici, Bhattacharjee, Caputa}, and Hilbert-space geometry~\cite{Pandey, Tiutiakina}. Furthermore, various diagnostics of quantum chaos are believed to be related to each other and form the ``web of diagnostics''~\cite{Bhattacharyya, Kudler-Flam}.

Among this set of approaches to quantum chaos, out-of-time ordered correlation function (OTOC) is probably the most prominent and useful diagnostic. Unlike other diagnostics, OTOC naturally generalizes the definitions of classical Lyapunov exponent and butterfly effect to a quantum case. The OTOC is defined as an expectation value of the squared commutator, coincides with the squared Poisson bracket and reflects the exponential divergence of trajectories in the semiclassical limit $\hbar \to 0$:
\beq \label{eq:Poisson}
C_{ij}(t) = -\Big\langle \big[ \hat{q}_i(t), \hat{p}_j(0) \big]^2 \Big\rangle \approx \hbar^2 \big\{ q_i(t), p_j(0) \big\}^2 = \hbar^2 \left| \frac{\pd q_i(t)}{\pd q_j(0)} \right|^2 \approx \hbar^2 \frac{\| \delta \mathbf{z}(t) \|^2}{\| \delta \mathbf{z}(0) \|^2} \approx \hbar^2 e^{2 \kappa t}. \eeq
Here, we assume that in the classical limit, the system is described by generalized coordinates $q_i$ and canonical momenta $p_i$, $i = 1, \cdots\!, N$ (so that $\mathbf{z} = \left(\mathbf{q}, \mathbf{p}\right)$), which become operators $\hat{q}_i$ and $\hat{p}_i$ upon quantization. The angle brackets $\langle \cdots \rangle$ denote the expectation value over a suitable initial ensemble, e.g., over the thermal one\footnote{In general, both classical and quantum Lyapunov exponents may differ in different points of the phase space, so the averaging over an initial ensemble is necessary to define a universal diagnostic of chaos, cf. Sec.~\ref{sec:energy-temperature}.}. So, using Eq.~\eqref{eq:Poisson}, we can introduce the quantum Lyapunov exponent similarly to its classical counterpart:
\beq \label{eq:quantum-Lyapunov}
\kappa_q \approx \frac{1}{2 t} \log\left[ \frac{1}{\hbar^2} \frac{1}{N^2} \sum_{i,j} C_{ij}(t) \right], \quad \text{as} \quad \frac{1}{\kappa_q} \ll t \ll \frac{1}{\kappa_q} \log \frac{1}{\hbar}, \eeq
and define ``quantum chaotic systems'' as systems with $\kappa_q > 0$. In the semiclassical limit, quantum and classical Lyapunov exponents are expected to coincide~\cite{Cotler-semiclassical, Jalabert}, $\kappa_{cl} \approx \kappa_q$ as $\hbar \to 0$. Note that in quantum chaotic systems, correlations between $q_i(t)$ and $p_j(0)$ are gradually lost in time~\cite{Maldacena-bound}, so we expect the OTOC to approach a constant value at the timescale $t_* \sim \frac{1}{\kappa} \log \frac{1}{\hbar}$ and thus restrict the times in~\eqref{eq:quantum-Lyapunov} to $t \ll t_*$. This timescale is called the scrambling time\footnote{This is essentially a simplified analog of the Ehrenfest time $t_E$, at which the semiclassical description of a chaotic system breaks down~\cite{Chirikov-88, Aleiner-96, Silvestrov, Berman, Zaslavsky}. In a quantized classicaly chaotic system, Ehrenfest time is proportional to $t_E \sim \frac{1}{K} \log \frac{1}{\hbar}$, where $K = \sum_i \kappa_i^+$ is the Kolmogorov-Sinai entropy and $\kappa_i^+$ are positive Lyapunov exponents.} and goes to infinity as $\hbar \to 0$. Furthermore, the scrambling time has a universal lower bound that is saturated for quantum theories on the black hole background or corresponding holographic duals~\cite{Maldacena-bound, Lashkari} and resolves the no-cloning paradox~\cite{Hayden, Sekino}.

Thus, due to the close relation to quantum chaos and scrambling, OTOCs have received a great attention from the high-energy and condensed-matter physics communities and have been estimated in a large variety of models. To the moment, such correlation functions were calculated in the Sachdev-Ye-Kitaev model and Jackiw-Teitelboim gravity~\cite{Maldacena-SYK, Kitaev, Sarosi, Rosenhaus, Trunin-SYK, Maldacena-JT}, three-dimensional black hole in anti-de Sitter space~\cite{Shenker-1, Shenker-2, Shenker-3, Roberts-1}, two-dimensional conformal field theories~\cite{Roberts-2, Fitzpatrick, Turiaci}, de Sitter space~\cite{Anninos, Aalsma}, weakly coupled matrix field theory~\cite{Stanford:phi-4, Grozdanov}, nonlinear sigma model~\cite{Swingle}, and many other quantum many-body systems~\cite{Michel, Shen, Bohrdt, Klug, Patel, Wang, Steinberg, Tikhanovskaya}. Moreover, OTOCs were experimentally measured with ion traps~\cite{Garttner} and nuclear magnetic resonanse platforms~\cite{Li} (see also a recent pedagogical review~\cite{Xu-tutorial}).

Nevertheless, there are few examples of systems where both classical and quantum Lyapunov exponents were calculated and compared~\cite{Hashimoto, Akutagawa, Hashimoto-2, Chavez-Carlos, Rozenbaum-1, Rozenbaum-2, Xu, Pilatowsky-Cameo}, so the putative relation between these exponents remains relatively poorly understood. Therefore, it is useful to consider another tractable model, where the correspondence between the classical and quantum butterfly effects can be directly checked. As an example of such a model, we propose the following simple vector mechanics:
\beq \label{eq:S}
S = \int dt \left[ \sum_{i=1}^N \left( \frac{1}{2} \dot{\phi}_i \dot{\phi}_i - \frac{m^2}{2} \phi_i \phi_i \right) - \frac{\lambda}{4 N} \sum_{i \neq j} \phi_i \phi_i \phi_j \phi_j \right], \eeq
where we assume the large-$N$ limit, $N \gg 1$, and introduce the 't Hooft coupling $\lambda$ for convenience\footnote{Note that we eliminate the true mass scale $M$ (i.e., masses of oscillators) via the rescaling $\phi_i \to \sqrt{M} \phi_i$, $t \to M t$, $\lambda \to M^2 \lambda$. Keeping in mind the similarity of the model~\eqref{eq:S} and its higher-dimensional analogs~\cite{Swingle}, we also call the frequency $m$ the $(0+1)$-dimensional ``mass''.}. In addition, we assume the system to be thermal with an inverse temperature\footnote{In the classical and large-$N$ limit, the temperature is approximately equal to the energy per degree of freedom, $\beta \sim N/E$, where $E$ is the total energy of the system. For details, see Sec.~\ref{sec:energy-temperature}.}~$\beta = 1/T$. In what follows, we will also employ the high-temperature limit $\beta m \ll 1$ and $\beta m \ll \lambda/m^3$, in which many calculations significantly simplify. Moreover, we will show that in this limit, both classical and quantum Lyapunov exponents are positive and approximately equal to each other. 

We emphasize that the model~\eqref{eq:S} does not contain self-interaction terms of the form $\sum_i \phi_i^4$. We exclude these interactions to break down the $O(N)$ symmetry and avoid the integrability. In fact, it is easy to see that the $O(N)$-symmetric model has exactly $N$ independent conserved quantities (energy $H$ and $N-1$ Casimir operators, $L_k^2 = \sum_{j=2}^{k+1} \sum_{i=1}^{j-1} L_{ij}$, where $k=1, \cdots\!, N-1$, $L_{ij} = \phi_i \pi_j - \pi_i \phi_j$ are the angular momenta and $\pi_i$ are the canonical momenta), so it is classicaly integrable. Moreover, this model straightforwardly reduces to a single quartic oscillator, whose equations of motion are explicitly integrated using elliptic functions. On the contrary, the deformed $O(N)$ model~\eqref{eq:S} possesses a much richer dynamics due to the sophisticated form of its nonlinear potential, as we will show in the main part of the paper.

The paper is organized as follows. In Sec.~\ref{sec:quantum-chaos}, we analytically calculate the leading contribution to the quantum Lyapunov exponent in the $O(N)$-symmetric and full nonsymmetric versions of the model~\eqref{eq:S}. In the former case, quantum Lyapunov exponent is zero, while in the latter case it is small but positive. Moreover, in the large-$N$ and high-temperature limit, this exponent acquires a relatively simple form, $\kappa_q \approx 1.3 \sqrt[4]{\lambda T}/N$. In Sec.~\ref{sec:classic}, we numerically calculate the classical Lyapunov exponent in the full model~\eqref{eq:S}. This exponent also approximately scales as $\kappa_{cl} \approx 1.3 \sqrt[4]{\lambda T}/N$ in the large-$N$ and high-temperature limit, which supports the correspondence between the quantum and classical chaotic behavior. In Sec.~\ref{sec:end}, we discuss the results and conclude. In addition, in Appendices~\ref{sec:technique}, \ref{sec:langrange}, and~\ref{sec:subleading-self-energy}, we discuss various technical details regarding the calculation of OTOCs in the augmented Schwinger-Keldysh diagrammatic technique.

In this paper, we assume the Plank constant $\hbar = 1$ and the Boltzmann constant $k_B = 1$ if not stated otherwise.

\section{Quantum chaos}
\label{sec:quantum-chaos}

In this section, we analyze the quantum chaotic behavior of the model~\eqref{eq:S} using the methods of $(0+1)$-dimensional quantum field theory. In other words, we consider the quartic terms as a small perturbation and sum the leading perturbative corrections to the OTOCs using the augmented Schwinger-Keldysh diagrammatic technique on the two-fold Keldysh contour~\cite{Aleiner, Haehl}. A brief introduction to this technique and derivation of the key identities are presented in Appendix~\ref{sec:technique}. For convenience, we explicitly separate the $O(N)$-symmetric and nonsymmetric interactions:
\beq S = \int dt \bigg[ \sum_{i=1}^N \left( \frac{1}{2} \dot{\phi}_i^2 - \frac{m^2}{2} \phi_i^2 \right) - \underbrace{\frac{\lambda}{4 N} \sum_{i,j=1}^N \phi_i^2 \phi_j^2}_\text{symmetric} + \underbrace{\frac{\lambda}{4 N} \sum_{i=1}^N \phi_i^4}_\text{nonsymmetric} \bigg]. \eeq
In the large-$N$ limit, nonsymmetric terms provide only the subleading ($1/N$ at most) corrections to correlation functions from the $O(N)$-symmetric model. Nevertheless, these terms are crucial for the development of the quantum chaotic behavior.

As a measure of quantum chaos, we consider the regularized\footnote{We employ the standard approach to OTOC regularization, i.e., uniformly smear the thermal distribution between the two commutators. In quantum mechanics, this regularization is not necessary because all correlation functions always remain finite. Furthermore, it is easy to show that the high-temperature behavior of OTOCs and quantum Lyapunov exponent does not depend on the regularization. However, we prefer to work with symmetrically regularized OTOCs because they have a clearer physical meaning~\cite{Liao,Romero-Bermudez}.} average square of the commutator of quantum fields $\phi_i$ and $\phi_j$:
\beq \label{eq:C-averaged}
C(t) = \frac{1}{N^2} \sum_{i,j=1}^N C_{ij}(t, t; 0, 0), \eeq
where
\beq 
\quad \label{eq:C}
\begin{aligned}
C_{ij}(t_1, t_2; t_3, t_4) &= \tr \left\{ \rho^{1/2} \left[ \phi_i(t_1), \phi_j(t_3) \right]^\dag \rho^{1/2} \left[ \phi_i(t_2), \phi_j(t_4) \right] \right\} \\ &= -\left\langle \phi_{i,uc}(t_1) \phi_{i,dc}(t_2) \phi_{j,uq}(t_3) \phi_{j,dq}(t_4) \right\rangle.
\end{aligned} \eeq
In the last line, we assume the thermal initial distribution with an inverse temperature $\beta$ and free Hamiltonian $H_0$, $\rho = e^{-\beta H_0}/\tr\left[e^{-\beta H_0}\right]$, turn on the coupling constant $\lambda$ adiabatically after the moment $t_0 < 0$, and rewrite the correlator in terms of the augmented Schwinger-Keldysh technique on the two-fold contour; see Appendix~\ref{sec:SK-basics} for a detailed derivation of Eq.~\eqref{eq:C} and Fig.~\ref{fig:contour} for the definition of the two-fold contour. In this notation, angle brackets denote the correlation function (in the interaction picture) ordered along the Keldysh contour, and indices ``$uc$'', ``$uq$'', ``$dc$'', and ``$dq$'' denote the classical and quantum components of the field on the upper and lower folds of the contour, respectively (see Fig.~\ref{fig:contour} and Eq.~\eqref{eq:cq}). At the tree level, the OTOC~\eqref{eq:C} simply reduces to the product of two retarded propagators:
\beq C_{ij}(t_1, t_2; t_3, t_4) = G_{0,ij}^R(t_1, t_3) G_{0,ij}^R(t_2, t_4) = \theta\left(t_{13}\right) \theta\left(t_{24}\right) \frac{\sin\left(m t_{13}\right)}{m} \frac{\sin\left(m t_{24}\right)}{m} \delta_{ij}, \eeq
where we denote $t_{ab} = t_a - t_b$ and substitute the explicit expression for the retarded propagator. Note that the regularized square of the commutator in the free theory is evidently non-chaotic: it oscillates with frequency $m$ and always remains finite. At the same time, in an interacting theory, this correlator receives loop corrections, which can sum into an exponentially growing expression. Such loop corrections are calculated using the Schwinger-Keldysh technique with eight vertices (Fig.~\ref{fig:vertices}) and four propagators that connect classical and quantum components on different folds of the Keldysh contour (all other two-point correlators are identically zero):
\begin{figure}[t]
    \center{\includegraphics[scale=0.37]{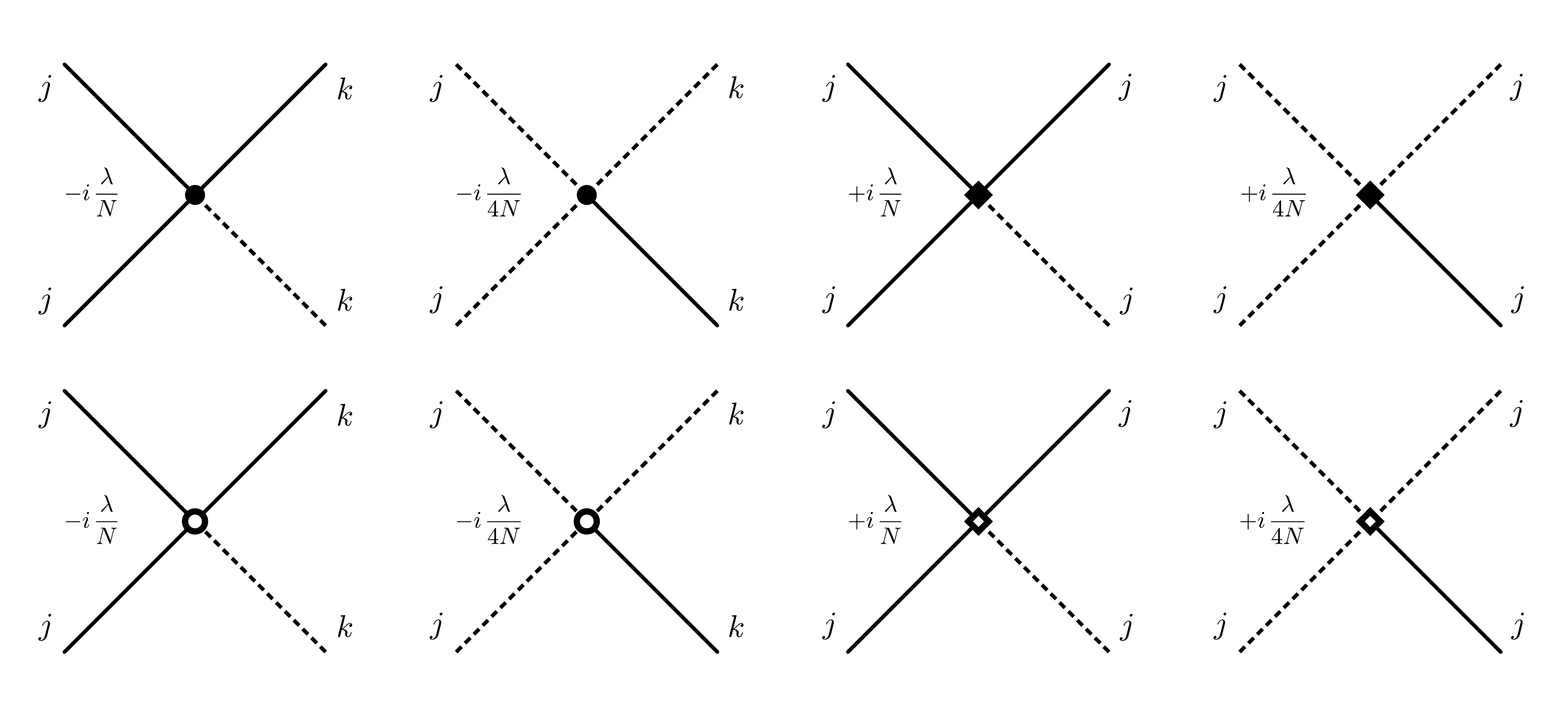}}
    \caption{Vertices and corresponding numerical factors in the augmented Schwinger-Keldysh diagrammatic technique of the model~\eqref{eq:S}. The solid and dashed lines correspond to the classical and quantum components; the black and white points correspond to vertices on the upper and lower folds of the Keldysh contour (Fig.~\ref{fig:contour}); the round and diamond points correspond to $O(N)$-symmetric and nonsymmetric vertices.}
    \label{fig:vertices}
\end{figure}
\beq \label{eq:propagators}
\begin{aligned}
i G_{ij}^R(t_1,t_2) &= \left\langle \phi_{i,uc}(t_1) \phi_{j,uq}(t_2) \right\rangle = \left\langle \phi_{i,dc}(t_1) \phi_{j,dq}(t_2) \right\rangle, \\
i G_{ij}^A(t_1,t_2) &= \left\langle \phi_{i,uq}(t_1) \phi_{j,uc}(t_2) \right\rangle = \left\langle \phi_{i,dq}(t_1) \phi_{j,dc}(t_2) \right\rangle, \\
i G_{ij}^K(t_1,t_2) &= \left\langle \phi_{i,uc}(t_1) \phi_{j,uc}(t_2) \right\rangle = \left\langle \phi_{i,dc}(t_1) \phi_{j,dc}(t_2) \right\rangle, \\
i G_{ij}^W(t_1,t_2) &= \left\langle \phi_{i,uc}(t_1) \phi_{j,dc}(t_2) \right\rangle= \left\langle \phi_{i,dc}(t_1) \phi_{j,uc}(t_2) \right\rangle.
\end{aligned} \eeq
At the tree level, these propagators have the following form (see Appendices~\ref{sec:SK-propagators} and~\ref{sec:Matsubara}):
\beq \label{eq:bare-propagators}
\begin{aligned}
&i G_{0;ij}^R(t_1, t_2) = i G_0^R(t_1, t_2) \delta_{ij}, &\quad &i G_0^R(t_1, t_2) = -i \theta(t_{12}) \, \frac{\sin \left( m t_{12} \right)}{m}, \\
&i G_{0;ij}^A(t_1, t_2) = i G_0^A(t_1, t_2) \delta_{ij}, &\quad &i G_0^A(t_1, t_2) = i \theta(-t_{12}) \, \frac{\sin \left( m t_{12} \right)}{m}, \\
&i G_{0;ij}^K(t_1, t_2) = i G_0^K(t_1, t_2) \delta_{ij}, &\quad &i G_0^K(t_1, t_2) = \frac{1}{2} \coth \frac{\beta m}{2} \, \frac{\cos \left( m t_{12} \right)}{m}, \\
&i G_{0;ij}^W(t_1, t_2) = i G_0^W(t_1, t_2) \delta_{ij}, &\quad &i G_0^W(t_1, t_2) = \frac{e^{\beta m/2}}{e^{\beta m} - 1} \, \frac{\cos \left( m t_{12} \right)}{m}.
\end{aligned} \eeq
Note that the retarded and advanced propagators are related by a simple permutation of their points: $G_{ij}^A(t_1, t_2) = G_{ij}^R(t_2, t_1)$. This relation straightforwardly follows from the definition of the propagators and holds both at the tree and loop levels.

Moreover, in the stationary situation and thermal equilibrium, all four real-time propagators~\eqref{eq:propagators} are unambiguously restored from a single imaginary-time (Matsubara) propagator using the analytic continuation procedure and the fluctuation-dissipation theorem, see Appendix~\ref{sec:Matsubara}. For simplicity, we restrict ourselves to such situations. However, the calculations in the Schwinger-Keldysh technique, including the calculations of this section, are easily extended to arbitrary initial states and nonstationary Hamiltonians, e.g., see~\cite{Trunin:ON, Akhmedov-1, Akhmedov-2, Arseev}.

Finally, note that the commutator~\eqref{eq:C-averaged} is not a direct analog of sensitivity from the classical version of the model~\eqref{eq:S}. Strictly speaking, the quantum Lyapunov exponent should be inferred from the following expectation value, which involves both the coordinate $\phi_j$ and the canonical momentum $\pi_i = \dot{\phi}_i$:
\beq \label{eq:C-averaged-correct}
c(t) = \frac{1}{N^2} \sum_{i,j=1}^N \pd_{t_1} \pd_{t_2} C_{ij}(t_1, t_2; 0, 0) \Big|_{t_1 = t_2 = t}. \eeq
However, if the commutator~\eqref{eq:C} grows exponentially before the averaging, $C_{ij} \sim e^{\kappa (t_1 + t_2 - t_3 - t_4)}$, both quantum Lyapunov exponents calculated from~\eqref{eq:C-averaged} and~\eqref{eq:C-averaged-correct} coincide with $\kappa$. Hence, these averaged correlators have the same qualitative behavior and both can be considered as a quantum analog of the Poisson bracket~\eqref{eq:Poisson} and classical sensitivity.

\subsection{Resummed propagators and vertices}

Let us sum the leading order, $\mO(1)$, loop corrections to propagators and vertices in the model~\eqref{eq:S} on the two-fold Keldysh contour. We remind that in this order, the nonsymmetric vertices are negligible, so the calculations in the full and $O(N)$-symmetric models approximately coincide.

First of all, consider loop corrections to propagators. To the leading order in $1/N$, these corrections are restricted to the so-called tadpole (or cactus) diagrams (Fig.~\ref{fig:tadpoles}), which simply shift the tree-level mass. This can be inferred directly from the system of Dyson-Schwinger equations on the $\mO(1)$ resummed propagators (we do not show the equation on the advanced propagator, which is easily reproduced from the identity $G^A(t_1,t_2) = G^R(t_2,t_1)$):
\begin{figure}[t]
    \center{\includegraphics[scale=0.37]{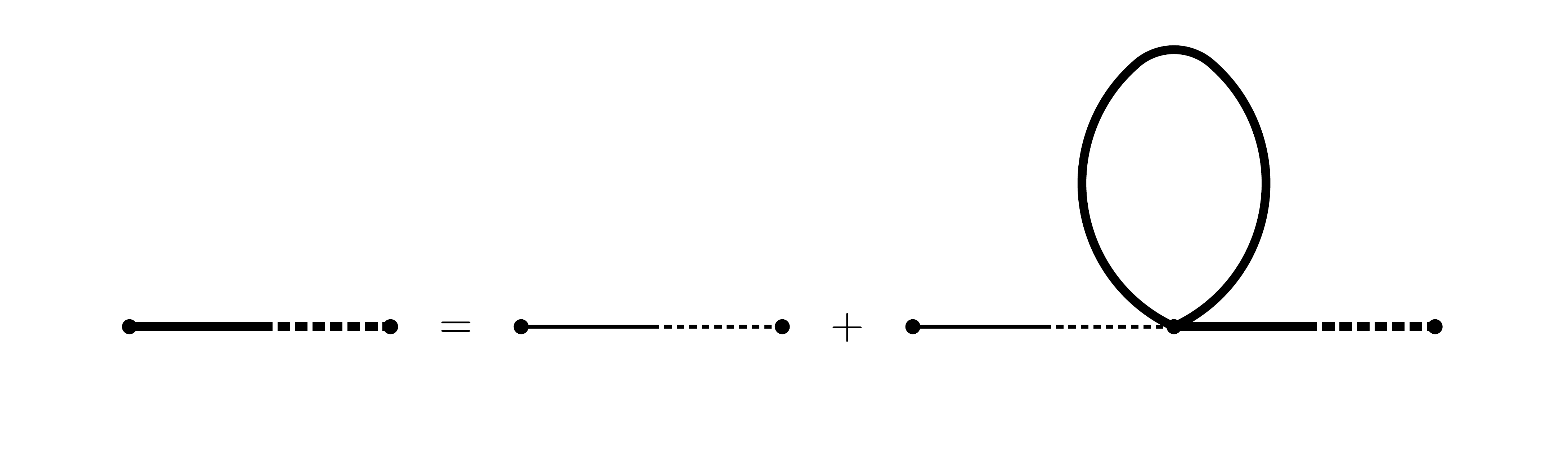}}
    \caption{An example of the Dyson-Schwinger equation on the propagators in the model~\eqref{eq:S}, which sums the leading order corrections to the retarded propagator. The thin and bold lines correspond to the tree-level and resummed propagators, respectively. The equations on the other three propagators have the same structure.}
    \label{fig:tadpoles}
\end{figure}
\beq \label{eq:tadpoles}
\begin{aligned}
G^R(t_1, t_2) &= G_0^R(t_1, t_2) + i \lambda \int_{t_0}^\infty dt \, G_0^R(t_1, t) G^K(t, t) G^R(t, t_2), \\
G^K(t_1, t_2) &= G_0^K(t_1, t_2) + i \lambda \int_{t_0}^\infty dt \left[ G_0^R(t_1, t) G^K(t, t) G^K(t, t_2) + G_0^K(t_1, t) G^K(t, t) G^A(t, t_2) \right], \\
G^W(t_1, t_2) &= G_0^W(t_1, t_2) + i \lambda \int_{t_0}^\infty dt \left[ G_0^R(t_1, t) G^K(t, t) G^W(t, t_2) + G_0^W(t_1, t) G^K(t, t) G^A(t, t_2) \right].
\end{aligned} \eeq
Applying the operator $\pd_{t_1}^2 + m^2$ to these equations and keeping in mind that the tree-level and resummed propagators have the same structure, we obtain the relation between the bare ($m$) and resummed ($\tilde{m}$) masses:
\beq \label{eq:thermal-m}
\frac{\tilde{m}^2}{m^2} = 1 + \frac{\lambda}{2 m^3} \frac{m}{\tilde{m}} \coth \left( \frac{\beta m}{2} \frac{\tilde{m}}{m} \right). \eeq
In general, this transcendental equation has a single real positive solution that depends on $\beta m$ and $\lambda/m^3$ in a complex way. However, we can approximately solve it in the low-temperature, $\beta m \gg 1$, or high-temperature, $\beta m \ll 1$ and $\beta m \ll \lambda/m^3$, limits. In the first case, the resummed mass is fully determined by the coupling constant; moreover, it approximately coincides with the bare mass when $\lambda/m^3 \ll 1$. In the second case, the resummed mass is proportional to the quartic root of the coupling constant and temperature, $\tilde{m} \approx \sqrt[4]{\lambda / \beta} \gg m$. Therefore, in the high-temperature limit, temperature sets the only reasonable energy scale (in this case, the resummed mass is usually called thermal). Roughly speaking, this happens because in this limit, the potential energy stored in nonlinear terms exceeds the energy stored in quadratic terms (see the discussion in Sec.~\ref{sec:end}).

Note that Eq.~\eqref{eq:thermal-m} also straightforwardly follows from the Dyson-Schwinger equation in the Matsubara technique:
\beq G(i \omega_n) = G_0(i \omega_n) - \frac{\lambda}{\beta} \sum_{\omega_k} G_0(i \omega_n) G(i \omega_k) G(i \omega_n), \eeq
hence,
\beq G^{-1}(i \omega_n) = G_0^{-1}(i \omega_n) + \frac{\lambda}{\beta} \sum_{\omega_k} G(i \omega_k). \eeq
Substituting the explicit form of the bare and resummed propagators, we reproduce the equation on the $\mO(1)$ resummed mass:
\beq \omega_n^2 + \tilde{m}^2 = \omega_n^2 + m^2 + \frac{\lambda}{\beta} \sum_{\omega_k} \frac{1}{\omega_k^2 + \tilde{m}^2} = \omega_n^2 + m^2 + \frac{\lambda}{2 \tilde{m}} \coth \frac{\beta \tilde{m}}{2}. \eeq
We remind that the correspondence between the real-time and imaginary-time propagators holds only for the thermal state (see Appendix~\ref{sec:Matsubara}).

We also emphasize that the state of the full interacting theory~\eqref{eq:S} is different from the initial thermal state, which is defined with respect to the non-interacting Hamiltonian. However, in the leading order in $1/N$, this difference manifests itself in a mere shift of the mass in the one-particle distribution function $n = \langle a_i^\dag a_i\rangle = 1/\left(e^{\beta \tilde{m}} - 1\right)$ (no sum). Hence, in the large-$N$ limit, the state of the full interacting theory approximately coincides with a true thermal state. From the quantum mechanical point of view, this means that in the large-$N$ limit, the energy levels of the full Hamiltonian are shifted when the interactions are adiabatically turned on, but the crossing of the adjacent levels is excluded, and the degeneracy of each level is approximately conserved\footnote{In fact, the degeneracy is lifted only in the $1/N$ order and only due to the nonsymmetric interactions.}.

Now let us sum the leading order corrections to the vertices, which are given by the bubble chain diagrams (Fig.~\ref{fig:chains}). The corresponding Dyson-Schwinger equation has the following form:
\begin{figure}[t]
    \center{\includegraphics[scale=0.37]{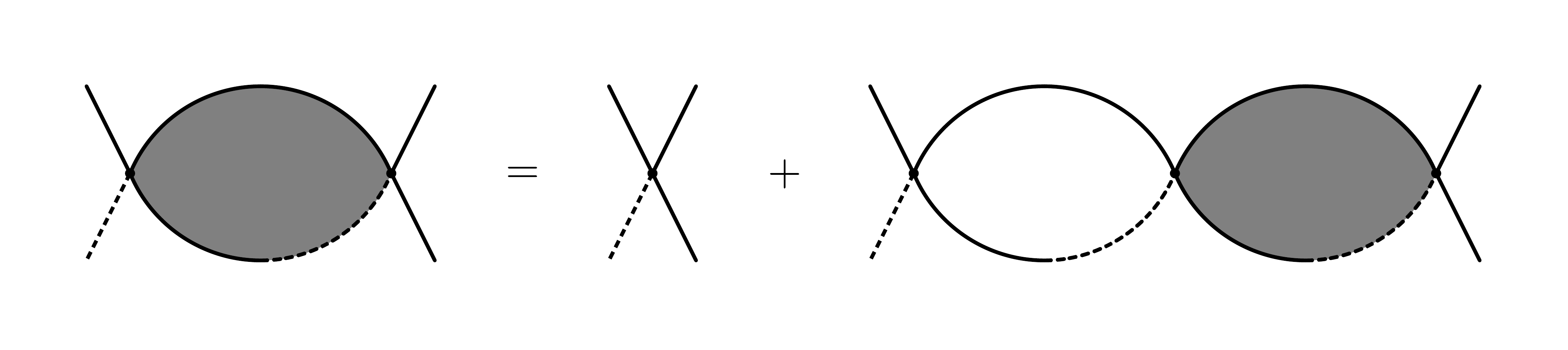}}
    \caption{An example of the Dyson-Schwinger equation that sums the leading order loop corrections to one of the vertices in the model~\eqref{eq:S}. The dark gray loop denotes the resummed ``bubble chain''. The equations that sum the leading order corrections to other vertices have the same structure.}
    \label{fig:chains}
\end{figure}
\beq \label{eq:bubble-chain}
B(t_1, t_2) = \delta(t_1 - t_2) + 2 i \lambda \int_{t_0}^\infty dt_3 \, G^R(t_1, t_3) G^K(t_1, t_3) B(t_3, t_2), \eeq
where $B$ denotes the resummed chain of $\mO(1)$ corrections to the vertex and $G^{K,R,A}$ denote the $\mO(1)$ resummed propagators~\eqref{eq:tadpoles}. For brevity, we suppress the external legs and the original numerical factor of the tree-level vertex (which corresponds to the Dirac delta function in Eq.~\eqref{eq:bubble-chain}).

We also note that after the summation, new types of vertices appear in the same order in $1/N$. However, all these vertices are derived from the resummed chain~\eqref{eq:bubble-chain} by adding a single appropriate bubble (compare with~\cite{Trunin:ON}). Moreover, such new vertices do not appear in the perturbative expansion of the resummed four-point correlator~\eqref{eq:C} due to causality reasons (see Secs.~\ref{sec:quantum-symmetric} and~\ref{sec:quantum-nonsymmetric}).

To solve Eq.~\eqref{eq:bubble-chain}, we propose the following ansatz, which is inspired by the structure of the single bubble:
\beq \label{eq:bubble-chain-result} B(t_1, t_2) = \delta(t_{12}) - \nu \, \tilde{m} \, \theta\left( t_{12} \right) \sin\left( \mu \tilde{m} t_{12} \right), \eeq
where $\mu$ and $\nu$ are real positive constants to be determined. Substituting this ansatz into Eq.~\eqref{eq:bubble-chain}, collecting the terms proportional to different oscillating functions and using the identity~\eqref{eq:thermal-m}, we obtain the following relations for $\mu$ and $\nu$:
\beq \mu^2 = 6 - 2 \, \frac{m^2}{\tilde{m}^2}, \qquad
\nu = \mu - \frac{4}{\mu}. \eeq
In the high-temperature limit, where the $\mO(1)$ resummed mass substantially exceeds the tree-level mass, $\tilde{m} \approx \sqrt[4]{\lambda/\beta} \gg m$, these relations are additionally simplified:
\beq \mu = \sqrt{6}, \qquad \nu = \sqrt{\frac{2}{3}}. \eeq
Thus, in this limit, the resummed bubble chain has the following form:
\beq B(t_1, t_2) = \delta(t_{12}) - \sqrt{\frac{2}{3}} \tilde{m} \theta\left( t_{12} \right) \sin\left( \sqrt{6} \tilde{m} t_{12} \right). \eeq
Note that essentially, the bubble chain~\eqref{eq:bubble-chain} is nothing but the resummed retarded propagator of the Lagrange field in the $O(N)$-symmetric model after the Hubbard-Stratonovich transformation (see Appendix~\ref{sec:langrange} for the details). However, we prefer to work with the original model~\eqref{eq:S} because it offers a clear setup for the calculation of real-time propagators.

First, it allows us to define the initial quantum state with respect to the free (Gaussian) Hamiltonian and then to turn on the coupling constant adiabatically, which is necessary to ensure the validity of Wick's theorem~\cite{Arseev, Hall, Kukharenko, Radovskaya}. In the theory after the Hubbard-Stratonovich transformation, this approach can potentially lead to a $0/0$ indeterminacy, so it should be used carefully.

Second, the calculations after the Hubbard-Stratonovich transformation require a careful specification of the initial quantum state of the Lagrange field. The explicit form of this state is obscure, since it forbids any classical-classical correlations but allows nontrivial classical-quantum ones. In particular, it cannot be a thermal state, so the fluctuation-dissipation theorem does not work for the propagators of the Lagrange field. In terms of the straightforward technique (Fig.~\ref{fig:vertices}), this means that there is no simple relation between the bubble chain~\eqref{eq:bubble-chain}, which fully lies on one fold, and the bubble chain that connects different folds. 

Finally, the full nonsymmetric model~\eqref{eq:S} cannot be rewritten using a single Lagrange field; in fact, one needs to introduce at least $N+1$ auxiliary fields to eliminate all quartic interaction terms. Thus, the calculations after the Hubbard-Stratonovich transformation become even less simple and transparent than before. Due to this reason, in the following subsections, we will employ the original diagrammatic technique (Fig.~\ref{fig:vertices}) with the $\mO(1)$ resummed propagators and vertices.

\subsection{Absence of quantum chaos in the \texorpdfstring{$O(N)$}{O(N)}-symmetric model}
\label{sec:quantum-symmetric}

In this subsection, we sum the leading corrections to the averaged correlator~\eqref{eq:C-averaged} and show that the resummed expression does not grow exponentially. To do this, we first consider the correlator~\eqref{eq:C} before the averaging. Keeping in mind the group structure of the vertices and propagators in the $O(N)$-symmetric version of the model~\eqref{eq:S}, we straightforwardly show that the most general expression for this correlator has the following form:
\beq C_{ij}(t_1, t_2; t_3, t_4) = \delta_{ij} F(t_1, t_2; t_3, t_4) + \frac{1}{N} H(t_1, t_2; t_3, t_4), \eeq
where we use the identities $\delta_{ij} \delta_{ij} = \delta_{ij}$ (no sum), $\delta_{ii} \delta_{jj} = 1$ (no sum) and introduce the factor $1/N$ for convenience. Substituting this expression into the Bethe-Salpeter equation on the four-point correlation function~\eqref{eq:C} and keeping only the leading (proportional to $1/N$) contribution to its kernel, we obtain the system of integral equations on $F$ and $H$:
\beq \begin{aligned}
F_{12;34} &= G_{13}^R G_{24}^R - \frac{8 \lambda^2}{N} \int dt_5 dt_6 dt_7 dt_8 \, G_{15}^R G_{26}^R B_{57} B_{68} G_{78}^W G_{78}^W \, F_{56;34}, \\ 
H_{12;34} &= - \frac{8 \lambda^2}{N} \int dt_5 dt_6 dt_7 dt_8 \, G_{15}^R G_{26}^R B_{57} B_{68} G_{78}^W G_{78}^W \, H_{56;34} \\ &\phantom{=} - \frac{4 \lambda^2}{N} \int dt_5 dt_6 dt_7 dt_8 \, G_{15}^R G_{26}^R G_{56}^W B_{57} B_{68} G_{78}^W \left( F_{78;34} + H_{78;34} \right),
\end{aligned} \eeq
where the integrations go from $t_0$ to infinity (however, the retarded propagators contain Heaviside theta-functions that narrow the integration limits) and we introduce a short notation for the arguments of functions, $f(t_1, t_2, \cdots, t_n) \equiv f_{1 2 \cdots n}$. Furthermore, we do not need to solve these equations separately if we want to estimate the averaged correlator. Indeed, after the averaging over all fields, we readily establish the identity
\beq C_{12;34} = \frac{1}{N} \left( F_{12;34} + H_{12;34} \right). \eeq 
Hence, the equation that sums the leading order corrections to the averaged correlator has the following form (this equation describes the so-called ladder diagrams, see Fig.~\ref{fig:ladders}):
\begin{figure}[t]
    \center{\includegraphics[scale=0.37]{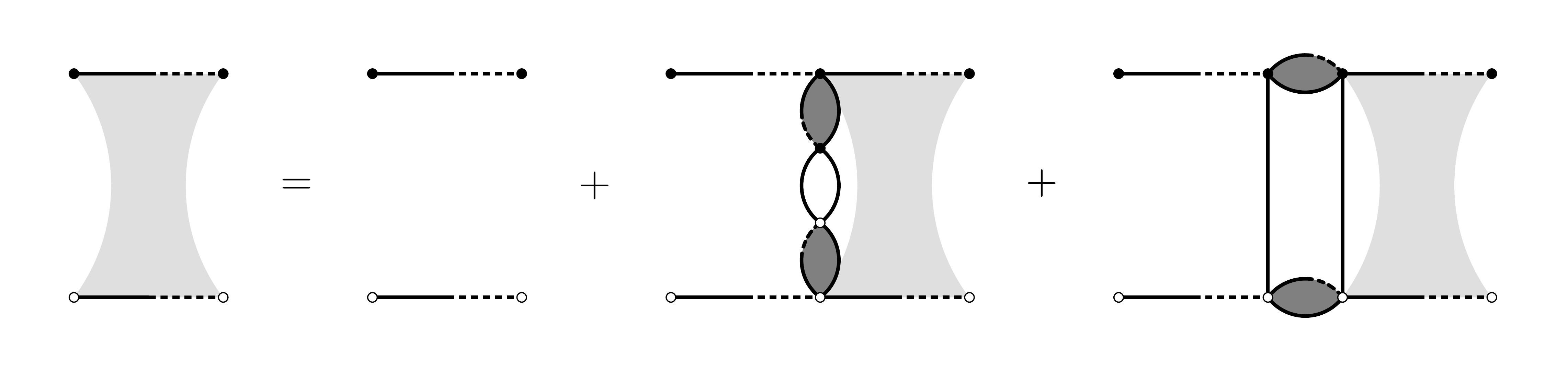}}
    \caption{A diagrammatic representation of the Bethe-Salpeter equation that sums the leading order corrections to the averaged correlator~\eqref{eq:C-averaged}. The lines denote the resummed tadpole diagrams from Fig.~\ref{fig:tadpoles}. The dark gray loops denote the resummed bubble chain diagrams from Fig.~\ref{fig:chains}. The light gray block denotes the resummed four-point correlator~\eqref{eq:C-averaged}.}
    \label{fig:ladders}
\end{figure}
\beq \label{eq:C-bethe-salpeter}
\begin{aligned}
C_{12;34} = \frac{1}{N} G_{13}^R G_{24}^R &- \frac{8 \lambda^2}{N} \int dt_5 dt_6 dt_7 dt_8 \, G_{15}^R G_{26}^R B_{57} B_{68} G_{78}^W G_{78}^W \, C_{56;34} \\ &- \frac{4 \lambda^2}{N} \int dt_5 dt_6 dt_7 dt_8 \, G_{15}^R G_{26}^R G_{56}^W B_{57} B_{68} G_{78}^W \, C_{78;34}.
\end{aligned} \eeq
Now, let us suppose that the resummed correlator exponentially grows with the average time, $C_{12;34} \sim e^{2 \kappa t}$, where $t = \frac{1}{2} \left(t_1 + t_2 - t_3 - t_4 \right)$. Substituting this ansatz into Eq.~\eqref{eq:C-bethe-salpeter}, taking all integrals and keeping only the leading exponentially growing terms, we establish the equation on the would-be Lyapunov exponent $\kappa$:
\beq \label{eq:k-sym}
1 \approx \frac{64}{N} \frac{w^2 \lambda^2}{\tilde{m}^6} \frac{1}{\mu^4} \frac{1}{\left( 1 + \frac{\kappa^2}{\tilde{m}^2} \right)^2} + \frac{4}{N} \frac{w^2 \lambda^2}{\tilde{m}^6} \frac{5 + \frac{\kappa^2}{\tilde{m}^2}}{\left( (\mu + 1)^2 + \frac{\kappa^2}{\tilde{m}^2} \right) \left( (\mu - 1)^2 + \frac{\kappa^2}{\tilde{m}^2} \right) \left( 1 + \frac{\kappa^2}{\tilde{m}^2} \right)}, \eeq
where we introduce a short notation for the dimensionless prefactor of the Wightman propagator, $w = e^{\beta \tilde{m}/2}/\left(e^{\beta \tilde{m}} - 1\right)$. In what follows, it is also convenient to combine this prefactor, coupling constant $\lambda$, and resummed mass $\tilde{m}$ into another dimensionless quantity $\alpha = w \lambda/\tilde{m}^3$. In the high temperature limit, this quantity is close to unity: $\alpha \approx 1$ when $\beta m \ll 1$ and $\beta m \ll \lambda / m^3$.

However, the solutions to Eq.~\eqref{eq:k-sym} are pure imaginary for arbitrary $\mu$ and $w$, i.e., for arbitrary temperatures and parameters of the model:
\beq \label{eq:k-sym-solved}
\begin{gathered}
\kappa_{1,2} \approx \pm i \tilde{m} \left( 1 + \frac{4}{\mu^2} \frac{\alpha}{\sqrt{N}} - \frac{4 \left( 3 \mu^2 - 8 \right)}{\mu^4 \left(\mu^2 - 4\right)} \frac{\alpha^2}{N} \right), \\
\kappa_{3,4} \approx \pm i \tilde{m} \left( \mu - 1 + \frac{2 \mu - \mu^2 + 4}{2 \mu^2 \left(2 - 3 \mu + \mu^2 \right)} \frac{\alpha^2}{N} \right), \\
\kappa_{5,6} \approx \pm i \tilde{m} \left( \mu + 1 + \frac{2 \mu + \mu^2 - 4}{2 \mu^2 \left(2 + 3 \mu + \mu^2 \right)} \frac{\alpha^2}{N} \right),
\end{gathered} \eeq
where we neglect the $\mO(1/N^2)$ terms that cannot be restored from the approximate Eq.~\eqref{eq:C-bethe-salpeter}. Furthermore, in the large-$N$ limit, the corrections to the right hand side of Eq.~\eqref{eq:k-sym} are suppressed by the powers of $1/N$; hence, these corrections cannot substantially affect the behavior of exponents~\eqref{eq:k-sym-solved}. In other words, solutions to the full Eq.~\eqref{eq:k-sym}, which contains all powers in $1/N$, are also pure imaginary in this limit. 

Thus, the resummed average correlator~\eqref{eq:C-averaged} in the $O(N)$-symmetric model cannot exponentially grow with time. Instead, it reduces to a sum of oscillating functions, with the leading contribution provided by oscillations at the frequency $\omega \approx \tilde{m}$. This is consistent with the integrability of the $O(N)$-symmetric model at the classical level.

\subsection{Quantum chaos in the full nonsymmetric model}
\label{sec:quantum-nonsymmetric}

However, the series of ladder diagrams (Fig.~\ref{fig:ladders}) is not the only series that contributes to the averaged correlator~\eqref{eq:C-averaged} in the full nonsymmetric model~\eqref{eq:S}. The leading nonsymmetric contribution to the kernel of the Bethe-Salpeter equation on $C_{12;34}$ is constructed from the diagrams from Fig.~\ref{fig:ladders} by replacing a single $O(N)$-symmetric vertex (round point) by the corresponding nonsymmetric one (diamond point from Fig.~\ref{fig:vertices}). In this section, we will show that summation of such diagrams results in an exponential growth of $C_{12;34}$, which indicates the quantum chaotic behavior of model~\eqref{eq:S}.

\begin{figure}[t]
    \center{\includegraphics[scale=0.6]{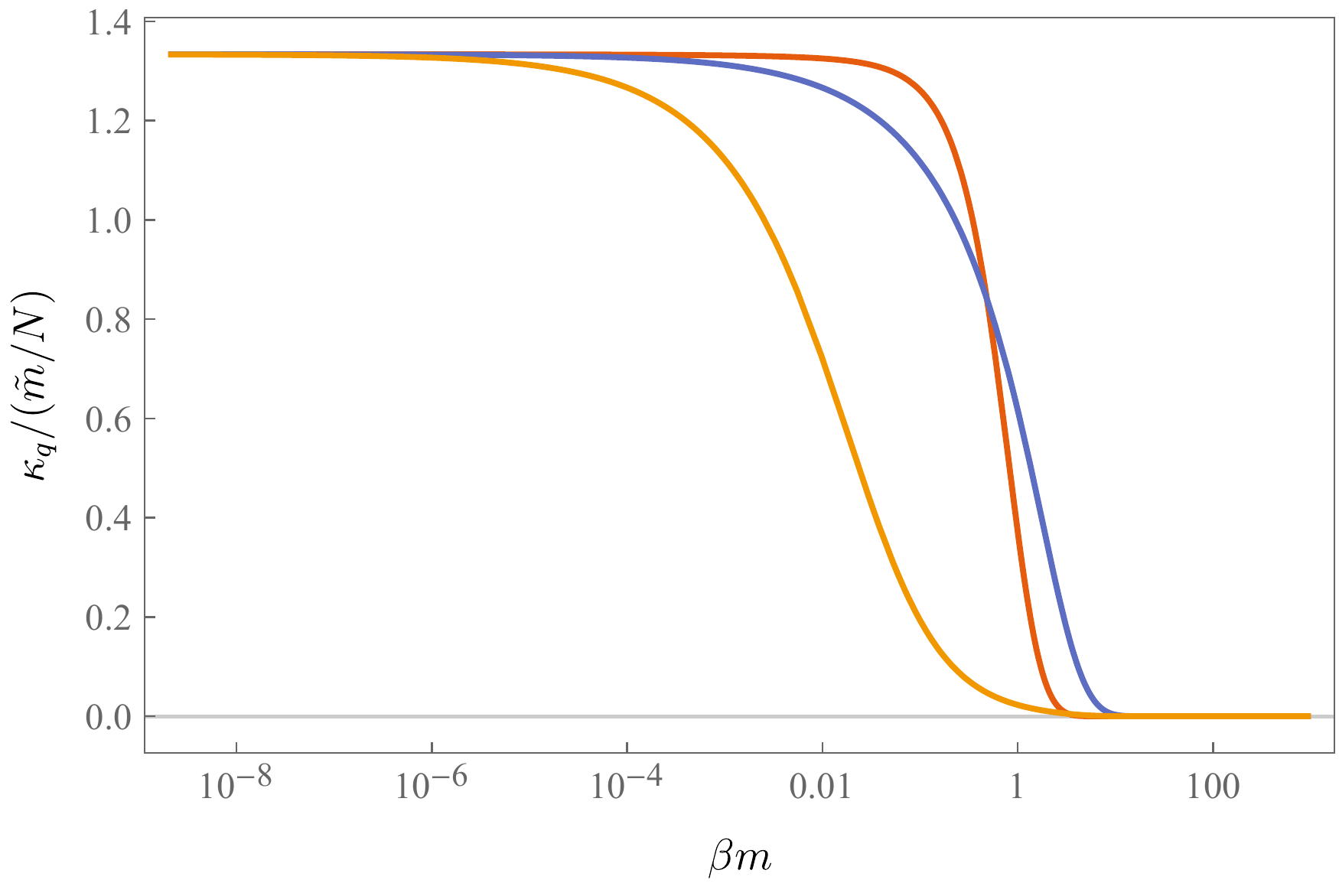}}
    \caption{An explicit temperature dependence of the quantum Lyapunov exponent $\kappa_q$ for different values of the 't Hooft coupling constant: $\lambda = 100$ (red), $\lambda = 1$ (blue), and $\lambda = 0.01$ (orange). For convenience, Lyapunov exponent is divided by $\tilde{m}/N$.}
    \label{fig:quantum-Lyapunov}
\end{figure}

Since two equivalent diagrams\footnote{That is, diagrams that have the same edge structure, but contain different number of symmetric and nonsymmetric vertices.} in the model~\eqref{eq:S} have essentially the same kinematics, the only difference between these diagrams is manifested in the total numerical factor. This factor consists of combinatorial factor, signs of vertices, and number of closed cycles (the latter determine the total power of $1/N$). For example, consider a bubble chain diagram with $n$ bubbles and $n+1$ symmetric vertices. Replacing one of its vertices with a nonsymmetric one, we gain an additional factor of $-3 (n+1)/N$, which consists of the combinatorial factor ($6(n + 1)/2$), different sign of the nonsymmetric vertex ($-1)$ and different number of cycles ($N^{n-1}/N^n$). Of course, these additional factors affect the summation of bubble chain diagrams, so the entire resummed bubble chain with one nonsymmetric vertex has a slightly different form comparing to~\eqref{eq:bubble-chain-result}:
\beq \begin{aligned}
\tilde{B}_{12} &= -\frac{3}{N} \int dt_3 \, B_{13} B_{32} \\ &= -\frac{3}{N} \delta_{12} + \frac{3}{N} \frac{\left( \mu^2 - 4 \right) \left( 3 \mu^2 + 4 \right)}{2 \mu^3} \theta\left( t_{12} \right) \tilde{m} \sin \left( \mu \tilde{m} t_{12} \right) + \frac{3}{N} \frac{\left( \mu^2 - 4 \right)^2}{2 \mu^2} \theta\left( t_{12} \right) t_{12} \cos \left( \mu \tilde{m} t_{12} \right).
\end{aligned} \eeq
Therefore, the equation that sums the leading nonsymmetric contributions to $C_{12;34}$ is slightly different from Eq.~\eqref{eq:C-bethe-salpeter}: 
\beq \label{eq:C-bethe-salpeter-nsymm}
\begin{aligned}
C_{12;34}^\text{nonsymm} = \frac{1}{N} G_{13}^R G_{24}^R &- \frac{8 \lambda^2}{N} \int dt_5 dt_6 dt_7 dt_8 \, G_{15}^R G_{26}^R \left( \tilde{B}_{57} B_{68} + B_{57} \tilde{B}_{68}\right) G_{78}^W G_{78}^W \, C_{56;34}^\text{nonsymm} \\ &- \frac{4 \lambda^2}{N} \int dt_5 dt_6 dt_7 dt_8 \, G_{15}^R G_{26}^R G_{56}^W \left( \tilde{B}_{57} B_{68} + B_{57} \tilde{B}_{68} \right) G_{78}^W \, C_{78;34}^\text{nonsymm}.
\end{aligned} \eeq
Similarly to the previous subsection, we search for exponentially growing solutions to this equation, $C_{12;34}^\text{nonsymm} \sim e^{2 \kappa t}$, $t = \frac{1}{2} \left( t_1 + t_2 - t_3 - t_4 \right)$. Substituting this ansatz into Eq.~\eqref{eq:C-bethe-salpeter-nsymm}, we obtain an analog of Eq.~\eqref{eq:k-sym}:
\beq \label{eq:k-nsym}
1 \approx -\frac{1536}{N^2} \frac{w^2 \lambda^2}{\tilde{m}^6} \frac{1}{\mu^6} \frac{1}{\left( 1 + \frac{\kappa^2}{\tilde{m}^2} \right)^2} - \frac{24}{N^2} \frac{w^2 \lambda^2}{\tilde{m}^6} \frac{\left( 5 + \frac{\kappa^2}{\tilde{m}^2} \right) \left( 3 \mu^2 - 3 + (\mu^2 + 6) \kappa^2 + \kappa^4 \right)}{\left( 1 + \frac{\kappa^2}{\tilde{m}^2} \right) \left( (\mu + 1)^2 + \frac{\kappa^2}{\tilde{m}^2} \right)^2 \left( (\mu - 1)^2 + \frac{\kappa^2}{\tilde{m}^2} \right)^2}, \eeq
where we again neglect the terms of higher orders in $1/N$. However, due to the negative sign of the right hand side, some solutions to this equation have a positive real part (we remind that for convenience, we introduced a dimensionless quantity $\alpha = w \lambda/\tilde{m}^3$):
\beq \begin{aligned}
\kappa_{1,2,3,4} &\approx \pm i \tilde{m} \pm \frac{8 \sqrt{6}}{\mu^3} \frac{\alpha}{N} \tilde{m}, \\
\kappa_{5,6,7,8} &\approx \pm i (\mu - 1) \tilde{m} \pm \sqrt{\frac{3 (\mu + 2) (2 \mu - \mu^2 + 4)}{4 \mu^3 (\mu - 1)}} \frac{\alpha}{N} \tilde{m}, \\
\kappa_{9,10,11,12} &\approx \pm i (\mu + 1) \tilde{m} \pm \sqrt{\frac{3 (\mu - 2) (2 \mu + \mu^2 - 4)}{4 \mu^3 (\mu + 1)}} \frac{\alpha}{N} \tilde{m},
\end{aligned} \eeq
Hence, the averaged commutator~\eqref{eq:C-averaged} does exponentially grow with time in the nonsymmetric model~\eqref{eq:S}. The maximal exponent of this growth --- the quantum Lyapunov exponent $\kappa_q$ --- is small but finite at finite $N$:
\beq \label{eq:lyap_quantum}
\kappa_q \approx \frac{8 \sqrt{6}}{\mu^3} \frac{\alpha \tilde{m}}{N}. \eeq
In the high-temperature limit, where $\mu \approx \sqrt{6}$ and $\alpha \approx 1$, this exponent scales as a quartic root of the temperature:
\beq \label{eq:lyap_high}
\kappa_q^\text{high} \approx \frac{4}{3} \, \frac{1}{N} \sqrt[4]{\frac{\lambda}{\beta}}. \eeq
In the low-temperature limit, the correlations between the different folds of the Keldysh contour are suppressed (see Eq.~\eqref{eq:bare-propagators}), so the maximal Lyapunov exponent is exponentially small in inverse temperature:
\beq \label{eq:lyap_low}
\kappa_q^\text{low} \approx \frac{\sqrt{6}}{N} \frac{\lambda}{m^3} m \exp\left(-\frac{\beta m}{2}\right), \eeq
where we additionally assume $\lambda/m^3 \ll 1$ for simplicity. At intermediate temperatures, the quantum Lyapunov exponent smoothly interpolates between these values, with smaller coupling constants corresponding to smoother transitions (Fig.~\ref{fig:quantum-Lyapunov}). The transition between the chaotic, Eq.~\eqref{eq:lyap_high}, and nonchaotic, Eq.~\eqref{eq:lyap_low}, behavior occurs at temperatures $\beta m \sim \min\left[1, \lambda/m^3 \right]$.

Finally, note that in higher-dimensional quantum field theories, one should also take into account thermalization processes that lead to an exponential damping of correlation functions and modify the quantum Lyapunov exponent (e.g., see~\cite{Stanford:phi-4, Grozdanov, Swingle}). However, we emphasize that this does not apply to the model~\eqref{eq:S}. In fact, the inverse dissipation time in this model is zero (or at least smaller than $\Gamma \sim \tilde{m}/N^2$) because this model has a finite number of degrees of freedom and thus cannot thermalize in a conventional sense, see Appendix~\ref{sec:subleading-self-energy}. Hence, the resummed correlator~\eqref{eq:C-averaged} does not suffer from an additional exponential damping, and approximation~\eqref{eq:lyap_quantum} is indeed $1/N$ exact.

\section{Classical chaos}
\label{sec:classic}

	In this section, we calculate the classical Lyapunov exponent of the model~\eqref{eq:S} and establish a qualitative correspondence with its quantum counterpart~\eqref{eq:lyap_high}.

	\subsection{General method}
	Before proceeding to a specific model, let us briefly discuss the notion of the classical Lyapunov exponent and provide the method for its calculation.
	
	Consider an arbitrary dynamical system defined by the Hamiltonian $H$ and the corresponding Hamilton's equations:
	\begin{equation}
		\begin{pmatrix}
			\dot x_i \\ \dot p_i
		\end{pmatrix}
		= \begin{pmatrix*}[c]
			\frac{\partial H}{\partial p_i} \\ -\frac{\partial H}{\partial x_i}
		\end{pmatrix*}.
	\end{equation}
	For convenience, we rewrite it in terms of the phase space coordinates $z_I = (x_i, p_i)$ as
	\begin{equation}
		\dot z_I = \pi_{IJ}\frac{\partial H}{\partial z_J}, \qquad 
		\pi = \begin{pmatrix}
			0 & 1 \\ -1 & 0
		\end{pmatrix},
	\end{equation}
	where $\pi_{IJ}$ has a meaning of the Poisson bi-vector in Darboux coordinates.
	
	As we discussed in the introduction, the classical Lyapunov exponent is broadly defined by the response to a perturbation of initial conditions, which grows exponentially with time. To define it more formally, consider a solution of the Hamilton's equations as a function of the initial condition $z_{0I}$, namely $z_I = z_I(t;z_0)$. We are interested in the large-time behavior of the sensitivity 
	\beq \Phi_{IJ}(t;z_0) = \frac{\partial z_I(t;z_0)}{\partial z_{0J}}, \eeq
	which represents the deviation of trajectories under small perturbations of initial conditions. Differentiating $\Phi$ in time and using Hamilton's equations, we get the following differential equation:
	\begin{align}
		\dot \Phi_{IJ} = \pi_{IK} \frac{\partial^2  H}{\partial z_K \partial z_L} \Phi_{LJ}.
	\end{align}
	with the initial condition $\Phi_{IJ}(0;z_0) = \delta_{IJ}$.
	The maximal singular value $\sigma_{\text{max}}(t)$ of the matrix $\Phi_{IJ}$ gives the maximal deviation of trajectories at a given time. Thus, expecting that $\sigma_{\text{max}}(t) \sim e^{\kappa t}$ for a chaotic system, we define the classical Lyapunov exponent as
	\begin{equation}
		\kappa = \lim_{t\to\infty} \frac1t \log \sigma_{\text{max}}(t). \label{cl_Lyap_def} 
	\end{equation}
	Hence, to find the classical Lyapunov exponent, one should solve the following system of ordinary differential equations on the variables $z_I$ and $\Phi_{IJ}$:
	\begin{align}
		&\dot z_I = \pi_{IJ}\frac{\partial H}{\partial z_J}, \label{Lyap_syst_1} \\
		&\dot \Phi_{IJ} = \pi_{IK} \frac{\partial^2  H}{\partial z_K \partial z_L} \Phi_{LJ}. \label{Lyap_syst_2}
	\end{align}
	
	\subsection{Relation to thermal Lyapunov exponent}
	\label{sec:energy-temperature}
	
	It is worth noting that Lyapunov exponent~\eqref{cl_Lyap_def} explicitly depends on initial conditions, so it cannot be considered a universal quantity. To eliminate this dependence and relate Lyapunov exponent to other universal quantities (e.g., temperature or total energy), it should be somehow averaged over the initial conditions. For example, we can consider a Gibbs ensemble with an inverse temperature $\beta$ or a microcanonical ensemble with a total energy $E$:
	\begin{equation}
		\langle\ldots \rangle_\beta =  \int d^{2N} z_0 \; e^{-\beta H(z_0)} (\ldots),  \qquad \langle\ldots \rangle_E =  \int d^{2N} z_0 \; \delta\left(H(z_0)-E\right) (\ldots),
	\end{equation}
	and define the classical Lyapunov exponent as an ensemble average of~\eqref{cl_Lyap_def}:
	\beq \label{eq:cl_Lyap_average}
	\overline{\kappa} = \lim_{t \to \infty} \frac{1}{t} \left\langle \log \sigma_\text{max}(t) \right\rangle. \eeq
	Note that we can also consider the following classical correlation function:
	\begin{equation} \label{eq:cl_exp_Lyap_average}
		C_{\text{cl}}(t) = \left\langle \Phi_{IJ}(t) \Phi_{JI}(t) \right\rangle, 
	\end{equation}
	which is nothing but the classical limit of the OTOC from Sec.~\ref{sec:quantum-chaos} in the Wigner quantization picture. Therefore, the Lyapunov exponent~$\kappa$ can be alternatively defined as the leading growth rate of $C_\text{cl}(t) \sim \exp(2\kappa t)$. However, this definition misses some important features of the classical chaos, so the definition~\eqref{eq:cl_Lyap_average} is preferable (see the discussion in Sec.~\ref{sec:end}).
	
	There is one more issue concerning the definition~\eqref{eq:cl_exp_Lyap_average} from the perspective of numerical computations. Suppose we have a sample of Lyapunov exponents $\{\kappa_n\}$ numerically calculated for a fixed energy or temperature but different initial conditions. Then, the numerical estimation of~\eqref{eq:cl_Lyap_average} gives the average Lyapunov exponent $\bar \kappa \sim \frac1n \sum_i \kappa_i$, whereas~\eqref{eq:cl_exp_Lyap_average} extracts the maximal one. Indeed, $C_{\text{cl}}(t) \sim \sum_i e^{2\kappa_i t}$, so $\kappa_\text{max} = \frac1{2t}\log C_{\text{cl}}(t) \to \max \kappa_i $ as $t \to \infty$. Both $\bar\kappa$ and $\kappa_\text{max}$ are expected to have the same order and qualitative behavior (cf. Figs.~\ref{fig:Lyap_N_dep}--\ref{fig:e_dep}), but the calculations of $\kappa_\text{max}$ usually require larger samples of initial conditions and produce larger errors due to a thin-tailed form of $\{ \kappa_n \}$ distribution. Due to this reason, we use the definition~\eqref{eq:cl_Lyap_average} to determine the general features of the classical Lyapunov exponent such as its dependence on~$E$ and~$N$.
	
	It may be difficult to calculate the thermodynamic Lyapunov exponent~\eqref{eq:cl_Lyap_average} even numerically. However, in some cases, it can be estimated by the microcanonical one. Namely, if the thermodynamic average saturates in a small vicinity of the phase-space subset that corresponds to a shell with a fixed energy $E$, then the microcanonical average with that energy will be close to the thermodynamic one, i.e., $\langle\ldots \rangle_E\simeq \langle\ldots \rangle_\beta$. More precisely, the condition for this approximate identity~is
	\beq \Delta E \ll \bar E, \eeq
	where
	\begin{equation}
		\bar E = \langle H \rangle_\beta, \qquad \Delta E^2 = \langle(H - \bar E)^2\rangle_\beta.
	\end{equation}
	As we will see in short, for the model~\eqref{eq:S}, this condition is satisfied in the large-$N$ limit.
	
	To find $\bar E$ and $\Delta E$, we evaluate the classical partition function $Z$ as
	\beq \begin{gathered}
	    Z(\beta) = \int d^N\phi \, d^N\pi \, \exp\Bigl[ - \beta H(\phi, \pi) \Bigr], \\
	    H(\phi, \pi) = \sum_{i} \left( \frac{1}{2} \pi_i^2 + \frac{m^2}{2} \phi_i^2 \right) + \frac{\lambda}{4 N} \sum_{i\ne j} \phi_i^2 \phi_j^2,
	\end{gathered} \eeq
	and use the standard formulas for energy averages $\bar E = - \partial_\beta \log Z(\beta)$, $\Delta E^2 = \partial_\beta^2 \log Z(\beta)$. Similarly to the quantum case (see Sec.~\ref{sec:quantum-chaos}), in the leading order in $1/N$, the full Hamiltonian $H(\phi, \pi)$ approximately coincides with the Hamiltonian of the $O(N)$-symmetric model. In turn, the partition function of the latter model can be easily calculated in the large-$N$ limit using the saddle point approximation in hyperspherical coordinates:
	\beq Z_{O(N)}(\beta) = \frac2{\Gamma(\frac{N}2)} \biggl[\frac{2N\pi^2m^2}{\beta\lambda}\biggr]^{N/2} \int_0^\infty d r \exp \biggl\{ -N \biggl[\frac{\beta m^3}\lambda \Bigl(\frac12 r^2 + \frac14 r^4 \Bigr) -\frac{N-1}{N} \log r \biggr] \biggr\}, \eeq
	\beq \log Z(\beta) \simeq C - \frac34 N \log \beta, \qquad \text{as} \qquad N \gg 1, \quad \beta \ll \frac{\lambda}{m^4}. \eeq
	Hence, 
	\begin{equation} \label{eq:energy_dep}
	    \bar E \simeq \frac34 N T, \qquad \Delta E^2 \simeq \frac34 N T^2, 
	\end{equation}
	where $T = \beta^{-1}$ is temperature and $\Delta E / \bar E = \mO\left(1/\sqrt{N}\right)$ goes to zero in the large-$N$ limit, as needed. Note that $E = \frac34 N T$ instead of naively expected $E = N T$. This is because the theory does not tend to any local field theory as $N \to \infty$, so its effective number of local degrees of freedom differs from $N$.

	\subsection{Classical Lyapunov exponents}
	
	To perform the numerical computations of the classical Lyapunov exponent in the model~\eqref{eq:S}, we first rescale the coordinate $x_i$, momentum $p_i$, time $\tilde{t}$, and energy $\tilde E$ using the corresponding dimensionful combinations of $m$ and $\lambda$:
	\begin{equation} \label{eq:dimless}
		[\phi] = \biggl[\frac{m}{\sqrt{\lambda}}\biggr], \qquad [\dot\phi] = \biggl[\frac{m^2}{\sqrt{\lambda}}\biggr], \qquad [t] = \biggl[\frac{1}{m}\biggr], \qquad [E] = \biggl[\frac{m^4}{\lambda}\biggr],
	\end{equation}
	so the action (\ref{eq:S}) reads
	\begin{equation}\label{eq:act_dimless}
		S = \frac{m^3}{\lambda}\int d\tilde t \, \bigg[ \sum_{i} \left( \frac{1}{2} \dot{x}_i^2 - \frac{1}{2} x_i^2 \right) - \frac{1}{4 N} \sum_{i\ne j} x_i^2 x_j^2\biggr].
	\end{equation}
	Assuming that for a fixed energy $E$, Lyapunov exponents weakly depend on initial conditions (this fact can be clarified from the plots), we expect the dimensionless Lyapunov exponent $\tilde \kappa$ to depend only on the number of degrees of freedom $N$ and the dimensionless energy $\tilde E$, i.e., $\tilde \kappa = \tilde \kappa_N(\tilde E)$. Therefore, the dimensionful Lyapunov exponent $\kappa$ can be recovered from the dimensionless $\tilde \kappa$ as
	\begin{equation}
		\kappa_N(E, \lambda, m) = m \, \tilde \kappa_N(\lambda E / m^4). \label{kappa_dim}
	\end{equation}
	To calculate $\tilde \kappa = \tilde \kappa_N(\tilde E)$, we numerically solve the system (\ref{Lyap_syst_1})--(\ref{Lyap_syst_2}) for the action~\eqref{eq:act_dimless}, which acquires the following explicit form:
	\begin{gather}
		\dot x_i = p_i, \qquad \dot p_i = - x_i - \frac1N x_i \sum_{j\ne i} x_j^2, \label{Lyap_syst_1_our}\\
		\dot \Phi = 
		\begin{pmatrix}
			0 & \delta_{kl}\\
			- \delta_{ij} - \frac{1}N \bigl(2 x_i x_j + \sum_{l}x_l^2 \, \delta_{ij} - 3 x_j^2 \, \delta_{ij}\bigr) & 0
		\end{pmatrix} \cdot \Phi. \label{Lyap_syst_2_our}
	\end{gather}
	With the aim of later comparison with the quantum Lyapunov exponent (\ref{eq:lyap_quantum}), we are mainly interested in two features of it's classical counterpart $\tilde \kappa$. The first is the dependence on the number of degrees of freedom for a fixed ``temperature'' (energy per degree of freedom), while the second is the dependence on $\tilde E$ for a fixed $N$.
	
	\begin{figure}[t]
	    \begin{center}
	    \includegraphics[width=\textwidth]{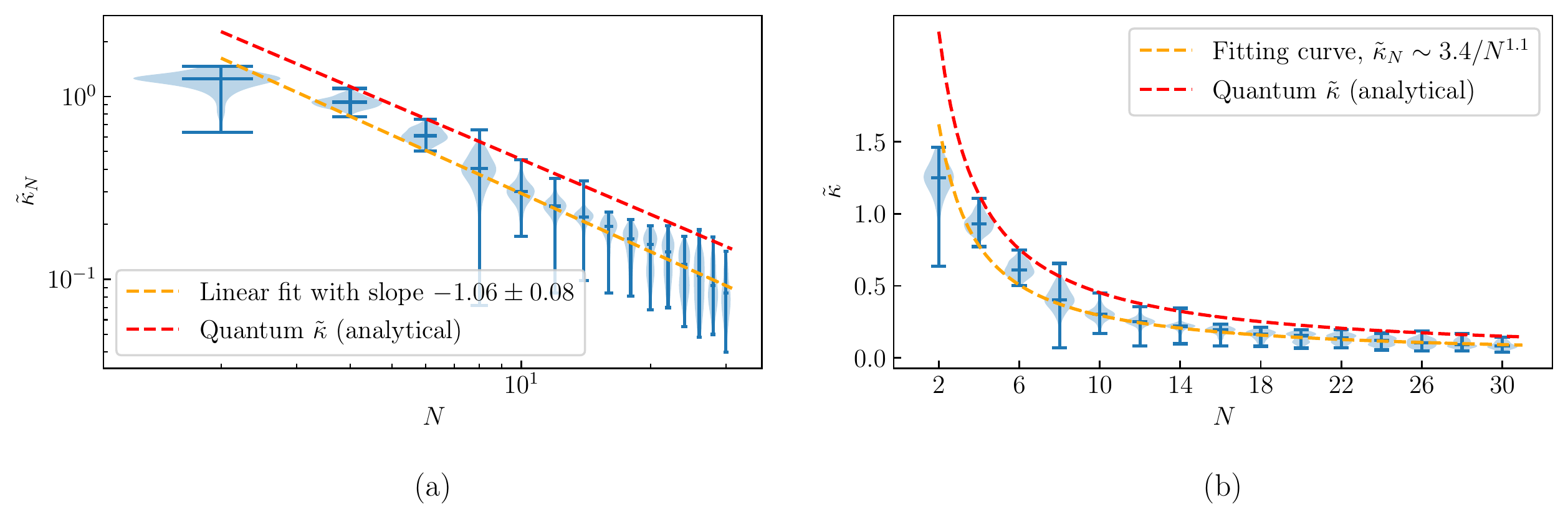}
	    \end{center}
	    \caption{Numerically calculated classical Lyapunov exponent $\tilde \kappa \sim 3.4 / N^{1.1}$ for $\tilde T = 100$. Vertical bars and violins represent the distribution of numerically calculated Lyapunov exponent for a fixed $N$ and energy $\tilde E = \tilde T N$. (a) is plotted in logarithmic scale and fitted by a linear curve to highlight the power-law dependence of $\tilde \kappa$ on $N$. (b)
	    is the same as (a), but in linear coordinates. For comparison, we also plot the analytically calcualted quantum Lyapunov exponent from Sec.~\ref{sec:quantum-chaos}. Note that this exponent approximately coincides with the upper edge of the classical distribution, i.e., with $\kappa_\text{max}$, as expected. We also emphasize that $\bar\kappa$ and $\kappa_\text{max}$ have approximately the same order and qualitative behavior.}
	    \label{fig:Lyap_N_dep}
	\end{figure}
	
	Let us begin with the first feature, namely, we fix $\tilde T = \tilde E / N$ and find the $\tilde \kappa$ for each $N$ in some range. Here, $\tilde T = \tilde E/N$ is the dimensionless temperature, up to the factor $4/3$ (cf. Eq.~(\ref{eq:energy_dep})). More specifically, our strategy is as follows:
	\begin{enumerate}
	    \item generate a huge number of initial conditions for each $N$ in some range, with the energy $\tilde E = \tilde T N$.
	    \item solve the system (\ref{Lyap_syst_1_our})--(\ref{Lyap_syst_2_our}) numerically for each initial condition generated\footnote{We use fourth-order Runge-Kutta method for the numerical integration. Symplectic integration schemes were not used, since (\ref{Lyap_syst_2_our}) is not of Hamiltonian form. Energy drift (energy non-conservation due to numerical integration artifacts) was controlled to be less than $1/1000$ of the initial energy.}.
	    \item find the Lyapunov exponent from numerical solution for $\Phi$ using the formula (\ref{cl_Lyap_def}) and average over initial conditions for each $N$.
	    \item fit the numerical results by a curve $\kappa_N(\tilde T N) = \varkappa/N^\gamma$, where $\varkappa$ and $\gamma$ are unknown\footnote{More accurately, we use the linear fit in the logarithmic coordinates: $\log \tilde{\kappa}_N = - \gamma \log N + \log\varkappa$, where $-\gamma$ is the slope and $\log \varkappa$ is the intercept.}.
	\end{enumerate}
	The result for $\tilde T = 100$ is presented in Fig.~\ref{fig:Lyap_N_dep} and shows that $\tilde\kappa_N(\tilde T N) \propto 1/N$, i.e., $\gamma \simeq 1$ with a good accuracy.
	
	\begin{figure}[t]
	    \centering
	    \includegraphics[width=0.7\textwidth]{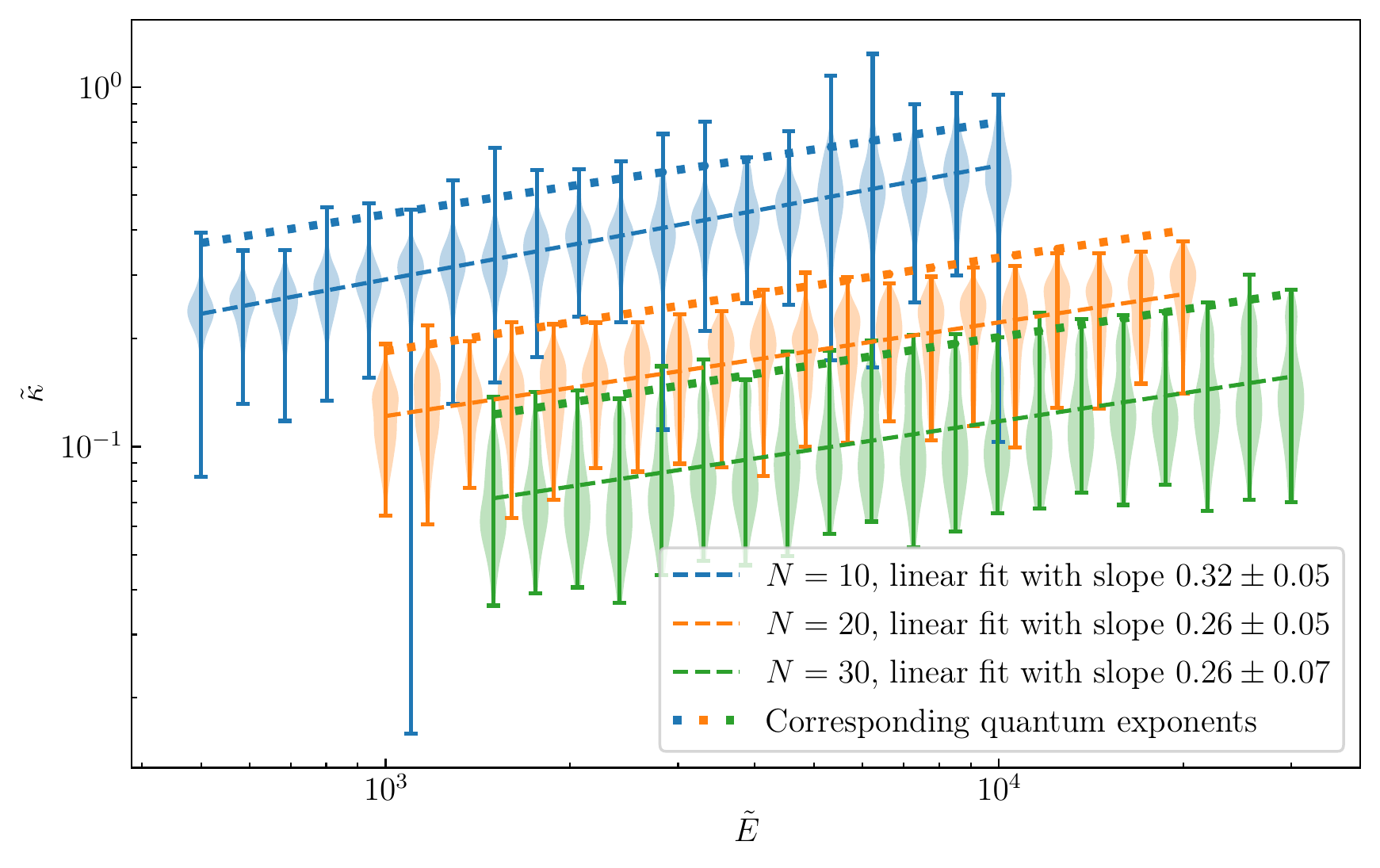}
	    \caption{Dependence of the numerically calculated classical Lyapunov exponent on energy for $N = 10$ (blue line), $N = 20$ (orange line), and $N=30$ (green line).
	    Thick dotted lines show the corresponding quantum Lyapunov exponents (\ref{eq:lyap_quantum}).
	    Vertical bars and violins represent the distribution of numerically calculated exponent for a fixed energy $\tilde E = \tilde T N$. The results are plotted in logarithmic scale and fitted by a linear curve to highlight the power-law dependence of~$\tilde \kappa$ on~$\tilde E$. For $N=20$ and $N=30$, the numerically calculated power-law dependence $\tilde \kappa_{cl} \propto \tilde E^{0.26}$ is close to the analytical one $\tilde \kappa_q \propto \tilde E^{1/4}$ (cf. Eq.~(\ref{eq:lyap_high})). For lower $N=10$, the discrepancy between the classical and quantum exponents is observed, though expected. Note that the quantum exponents approximately coincide with the upper edges of the corresponding classical distributions, i.e., with $\kappa_\text{max}$, as expected. We also emphasize that $\bar\kappa$ and $\kappa_\text{max}$ have approximately the same order and qualitative behavior in all three cases.}
	    \label{fig:e_dep}
	\end{figure}
	
	Similarly, we examine the energy dependence of the classical Lyapunov exponent for a fixed number of degreeds of freedom $N$. The technique of the numerical experiment is similar to the previous one and consists from the following steps. First, we generate a huge number of initial conditions corresponding to some energy $\tilde E$ and numerically calculate $\tilde \kappa$ for each of them. Then, we repeat the procedure for energies in some range. Expecting the power law dependence on $\tilde E$ (cf. Eq.~\eqref{eq:lyap_high}), we fit the numerical results by a curve $\tilde \kappa = \varkappa' \tilde E^{\gamma'}$ with coefficients $\varkappa'$ and $\gamma'$ to be determined. The results are shown in Fig.~\ref{fig:e_dep} and agree with the corresponding dependence of quantum Lyapunov exponent (\ref{eq:lyap_high}), i.e., $\tilde \kappa \propto \tilde E^{1/4}$. A slight deviation from this power-law dependence will be explained below.
	
	Finally, inspired by a good qualitative coincidence of the quantum and classical Lyapunov exponents, we propose a numerical experiment that determines the power-law dependence on $\tilde T$ (or corresponding energy $\tilde E = N \tilde T$) and $N$ simultaneously. Namely, we take the following ansatz:
	\begin{equation}
	    \tilde{\kappa} = \tilde\varkappa \,\tilde T^{\gamma_1} / N^{\gamma_2}, \quad \text{i.e.}, \quad \log \tilde\kappa = \gamma_1 \log \tilde T - \gamma_2 \log N + \log \tilde \varkappa,
	\end{equation}
	and determine $\gamma_1$, $\gamma_2$, and $\tilde \varkappa$ using linear regression for the Lyapunov exponents numerically calculated for a huge number of pairs $(\tilde T, N)$. We use the described procedure for the same range of $\tilde E$ and $N$ as in Figs.~\ref{fig:Lyap_N_dep}--\ref{fig:e_dep} and find the following estimates for the unknown coefficients:
	\begin{equation}
	    \gamma_1 = 0.28 \pm 0.02, \qquad \gamma_2 = 1.18 \pm 0.05, \qquad \tilde \varkappa = 1.29 \pm 0.22, \label{eq:3d_linregress}
	\end{equation}
	so we observe a slight deviation from the expected power-law dependence (\ref{eq:lyap_high}). There are two reasons for this mismatch. The first reason is the finite energy corrections to the power-law (\ref{eq:lyap_high}). Indeed, Eq.~(\ref{eq:lyap_high}) is obtained from Eq.~(\ref{eq:lyap_quantum}) in the high-temperature (i.e., high-energy) limit. For the classical Lyapunov exponent, the same behavior is expected in the same limit for the dimensional reasons (see the discussion in Sec.~\ref{sec:end}). We can partially control this finite energy effect by expanding (\ref{eq:lyap_quantum}) in the powers of $\beta$ (i.e., inverse powers of energy) in the high-temperature and semiclassical limit\footnote{Note that the last term in the square brackets is nothing but the leading genuine quantum correction to the classical Lyapunov exponent. This contribution has the same power in $\beta$ as the previous term in the expansion (so these terms can be confused if one sets $\hbar = 1$), but clearly vanishes in the semiclassical limit $\hbar \to 0$. In contrast, the preceding terms of the expansion are the purely classical finite temperature corrections to (\ref{eq:lyap_high}).}:
	\begin{equation} \label{eq:classical-expansion}
	    \kappa = \frac{4}{3} \frac{1}{N} \left(\frac{\lambda}{\beta}\right)^{\frac14}\! \left[1 - \frac{1}{4} \sqrt{\frac{\beta m^4}{\lambda}} - \frac{13}{96} \frac{\beta m^4}{\lambda} + \frac{91}{3456} \left(\frac{\beta m^4}{\lambda}\right)^{\frac32} - \frac{1}{16} \hbar^2 \lambda^{\frac12} \beta^{\frac32} + \ldots\right].
	\end{equation}
	If we try to fit the energy dependence~\eqref{eq:classical-expansion} with a simple power law for relatively small energies, we will get a power slightly exceeding the expected $\gamma_1 = 0.25$. The correct power is reproduced only for sufficiently large energies, where the subleading corrections are negligible. Unfortunately, such energies are elusive for usual numerical integration schemes in the large $N$ limit.
	
	The second reason for the mismatch of (\ref{eq:lyap_high}) and the numerical results (\ref{eq:3d_linregress}) is the finite $N$ corrections. Indeed, the result (\ref{eq:lyap_high}) was obtained in the leading order in $1/N$ expansion, so it might have the subleading correction terms. For the values of $N$ used in the numerical experiment ($N \sim 10$--$30$), the order of the finite $N$ correction is expected to be about ten percent, which is comparable to the deviation of $\gamma_2$ from unity. The impact of finite $N$ correction can be seen, for instance, on Fig.~\ref{fig:e_dep}, where we found a good coincidence of the power-law dependence on energy $\tilde\kappa \sim \tilde E^{1/4}$ for $N=20$ and $N=30$, whereas for the smaller value of $N$, namely $N=10$, we observe a slight deviation from this behavior.
	
	
\subsection{Analogy to billiards}
\label{sec:billiards}

\begin{figure}[t]
\begin{subfigure}{.5\textwidth}
  \centering
  \includegraphics[width=.9\linewidth]{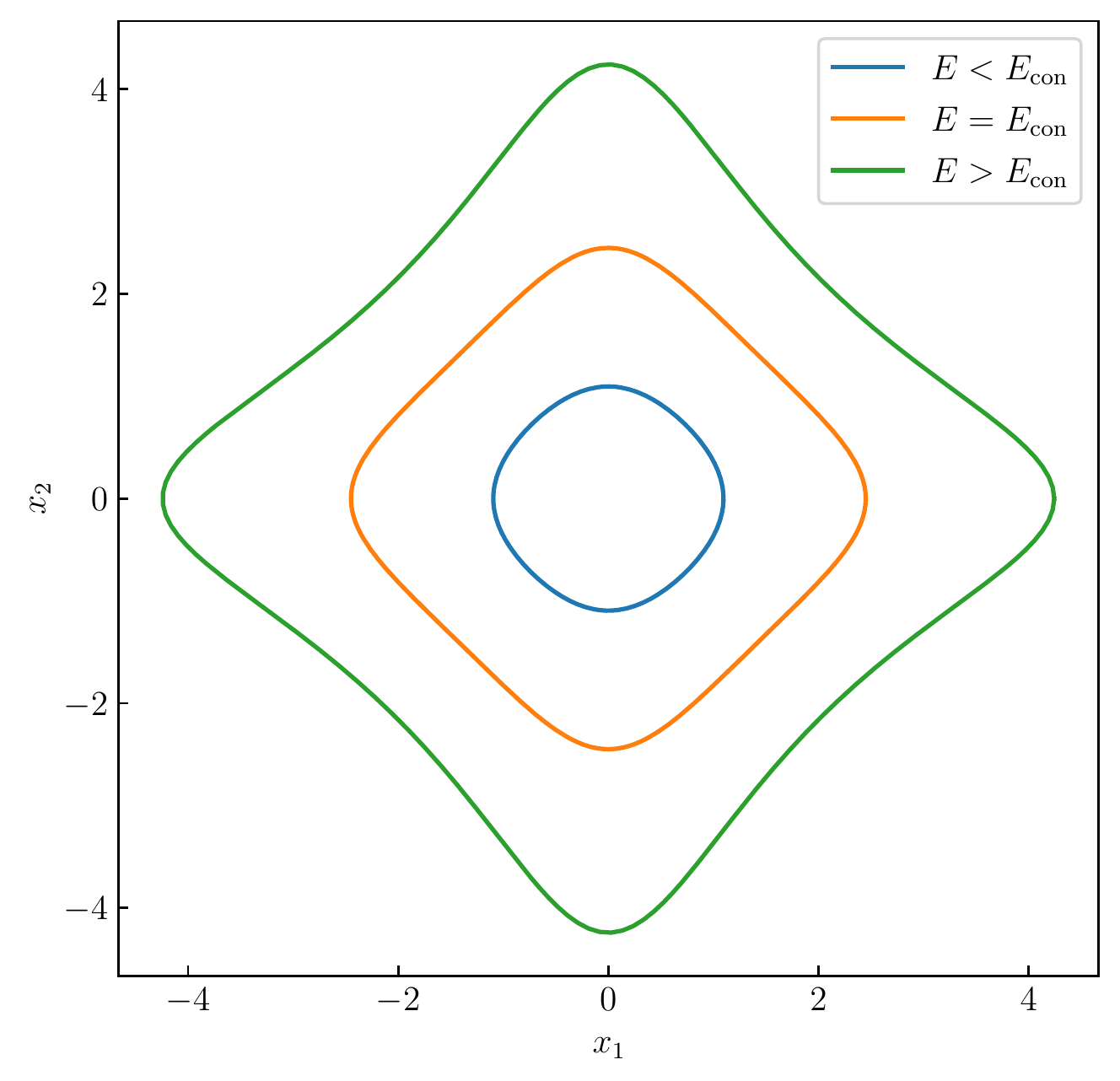}
  \caption{}
\end{subfigure}%
\begin{subfigure}{.5\textwidth}
  \centering
  \includegraphics[width=0.9\linewidth]{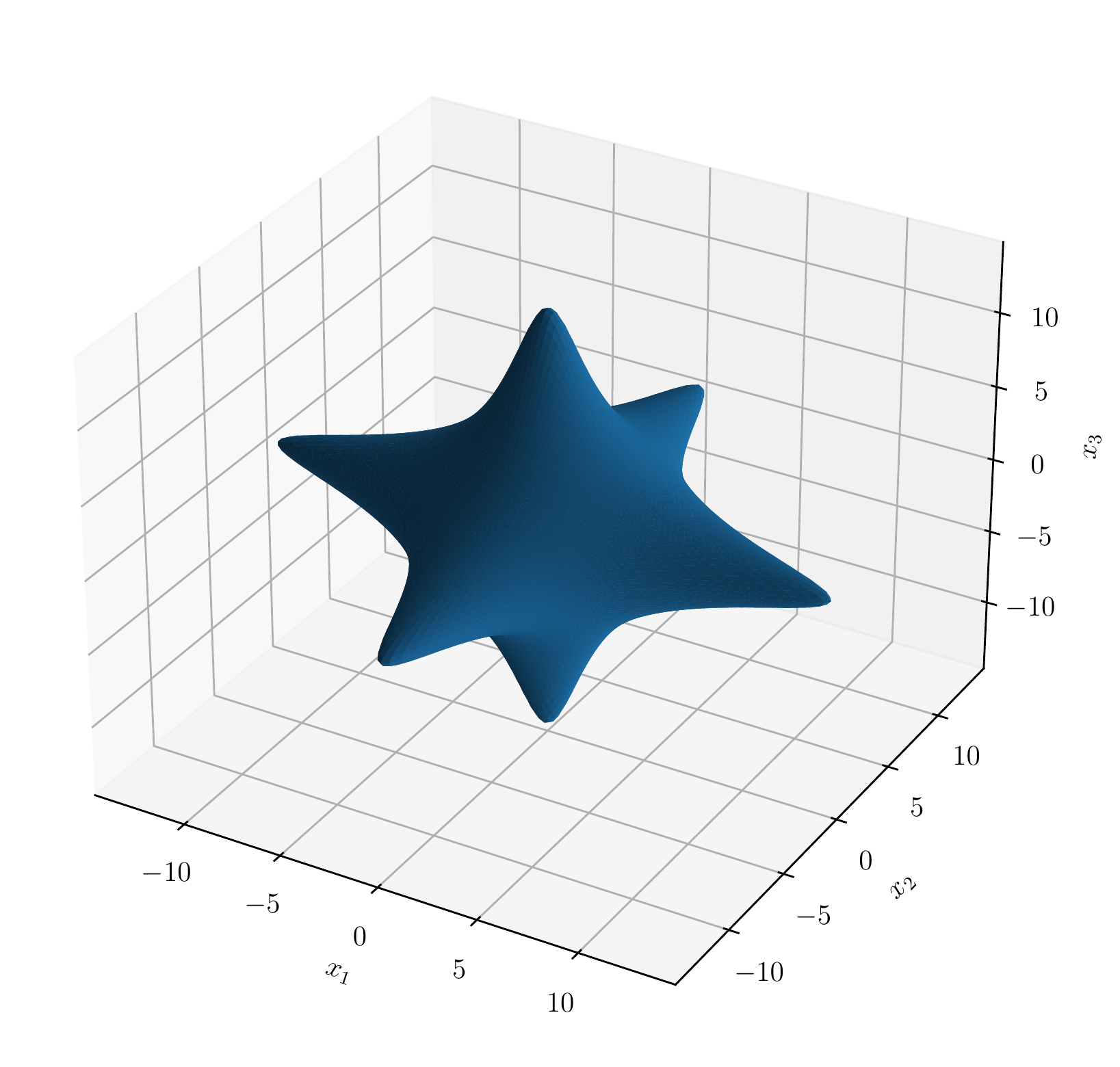}
  \caption{}
\end{subfigure}
\caption{(a) constant (potential) energy curve for $N=2$ and $E<E_{\text{con}}$ (blue line), $E=E_{\text{con}}$ (orange line), $E>E_{\text{con}}$ (green line). (b) constant energy surface for $N=3$ and $E \gg E_{\text{con}}$. All the quantities are dimensionless (cf. (\ref{eq:dimless})--(\ref{eq:act_dimless})).}
\label{fig:energy_shell}
\end{figure}

In addition to quantitative signatures of chaos, one may wonder the qualitative one. The qualitative argument is as follows. First of all, let us discuss regions of constant energy in the phase space. For vanishing momenta, i.e., for turning points, surfaces of constant energy have the topology of $S^{N-1}$. However, the properties of these surfaces are different for high and low energies. For low energies, the quartic term in the potential is subdominant, and the surface has the form of a slightly deformed Euclidean sphere embedded in $N$-dimensional configurational space. On the contrary, for high energies, constant energy surface has the form of an $N$-dimensional ``morning star'' (see Fig.~\ref{fig:energy_shell}). The distinctive feature of the latter is the presence of concave regions that appear at energies $E > E_{\text{con}} = 3 N m^4 / 2\lambda$. Therefore, we can make an analogy to Sinai billiards that exhibit a chaotic behavior in the presence of concave walls (or convex obstacles), see~\cite{Sinai-1, Sinai-2, Chernov-1, Chernov-2}. In our case, the role of walls is played by surfaces of constant energy that become concave for high energies. Note that in the $N=2$ version of the model~\eqref{eq:S}, chaotic behavior disappears exactly at such energies, compare with~\cite{Akutagawa}.

\section{Discussion and Conclusion}
\label{sec:end}

In this paper, we considered classical and quantum butterfly effects, i.e., calculated classical and quantum Lyapunov exponents in a simple vector model~\eqref{eq:S}. In the quantum case, we resummed the leading nonsymmetric contributions to the average square of the commutator~\eqref{eq:C-averaged} using the augmented Schwinger-Keldysh technique on a symmetric two-fold Keldysh contour and assuming the large-$N$ limit. We established the exponential growth of~\eqref{eq:C-averaged} at early periods of evolution, where this correlator is far from the saturation and ladder diagrams dominate its perturbative expansion. The quantum Lyapunov exponent, which determines the early-time growth of~\eqref{eq:C-averaged}, is approximately equal to $\kappa_q \approx 1.3 \sqrt[4]{\lambda T}/N$ in the high-temperature limit ($\beta m \ll 1$ and $\beta m \ll \lambda/m^3$) and exponentially suppressed, $\kappa_q \sim e^{-\beta m}/N$, in the low-temperature limit ($\beta m \gg 1$). In the classical case, we calculated the Lyapunov exponent numerically solving the equations of motion for a fixed temperature and number of degrees of freedom in some range. As a result, we established the high-temperature behavior $\kappa_{cl} \approx (1.3 \pm 0.2) (\lambda T)^{0.28 \pm 0.02}/N^{1.18 \pm 0.05}$, which qualitatively coincides with the behavior of quantum Lyapunov exponent. This coincidence supports the use of OTOCs as a diagnostic of quantum chaos in quantum many-body systems with a large number of degrees of freedom\footnote{In our model, the high-temperature limit is essentially the semiclassical limit, which is easy to see after the restoration of dimensional constants: $\hbar m \ll k_B T \sim E/N$.}. 

In fact, the qualitative energy dependence of Lyapunov exponents is easily restored from dimensional grounds~\cite{Akutagawa, Chirikov}. On one hand, for high energies, we can neglect the quadratic term in the Hamiltonian of the model~\eqref{eq:S}:
\beq H^\text{high} \approx \sum_{i=1}^N \frac{1}{2} \pi_i^2 + \frac{\lambda}{4 N} \sum_{i \neq j} \phi_i^2 \phi_j^2. \eeq
This pruned Hamiltonian is invariant under the following scale transformations:
\beq t \to \alpha^{-1} t, \qquad \phi_i \to \alpha \phi_i, \qquad H \to \alpha^4 H, \eeq
with an arbitrary positive constant $\alpha$. Since the Lyapunov exponent has the dimension of inverse time, this invariance implies the high-temperature dependence $\kappa \sim \sqrt[4]{E}$. On the other hand, for energies smaller than $E \sim N \hbar m$, the quartic interaction term is negligible, so the system becomes approximately free. From the classical point of view, positive Lyapunov exponents vanish for such small energies due to the Kolmogorov-Arnold-Moser theorem~\cite{Kolmogorov, Arnold, Moser}. From the quantum mechanical point of view, Lyapunov exponents are also heavily suppressed in this limit because it implies the vanishing correlations between different folds of the Keldysh contour (cf. Eq.~\eqref{eq:bare-propagators}).

We emphasize that OTOCs should be used with caution in systems with unstable fixed points, e.g., see~\cite{Rozenbaum-1, Rozenbaum-2, Xu, Pilatowsky-Cameo, Hashimoto-2}. In such systems, the quantum Lyapunov exponent is determined by the exponential divergence of trajectories near the unstable points, which is not necessarily imply chaos. This is a consequence of a slight difference in the definition of classical and quantum Lyapunov exponents: the $\kappa_{cl}$ is defined as the phase space average of the log of sensitivity, whereas the $\kappa_q$ is defined as the log of the phase space average of sensitivity. Nevertheless, the potential of the model~\eqref{eq:S} does not have any unstable fixed points --- as we pointed out in Sec.~\ref{sec:billiards}, it rather models an $N$-dimensional Sinai billiard with soft concave walls. Hence, the OTOCs correctly describe the chaotic behavior of this system.

The analysis of this paper can be extended in several possible directions. First, it is interesting to calculate the OTOCs for non-thermal initial states, e.g., the eigenstates of the free Hamiltonian or the coherent states that correspond to some classical solutions in the model~\eqref{eq:S}. In particular, the analytical calculations in the latter case might shed an additional light on the correspondence between the classical and quantum Lyapunov exponents. Besides, it is interesting to study the relationship between the scrambling and delocalization of coherent states~\cite{Brickmann, Gutschick}.

Second, it is promising to study other diagnostics of quantum chaos in our system, e.g., the Lanczos coefficients and Krylov complexity~\cite{Parker, Avdoshkin, Gorsky, Smolkin, Rabinovici, Bhattacharjee, Caputa}. Due to simplicity of the model~\eqref{eq:S}, these quantities should also be amenable to analytical calculations. For this reason, we expect this convenient example to help us understand the relationship between different diagnostics of quantum chaos. Moreover, such a relationship would provide us with an extra toolkit. In particular, it is interesting to check how the geometric approach of~\cite{Caputa}, which was developed for the calculation of Krylov complexity, explains the emergence of quantum chaos after the breaking of the continuous $O(N)$ symmetry down to the discrete group of symmetries of the model~\eqref{eq:S}.

Third, the developed method may be applied to various \textit{nonstationary} quantum systems, e.g., the generalization of the model~\eqref{eq:S} with an external force $f(t)$ and Markovian dissipation:
\beq \begin{gathered}
H = \frac{1}{2} \sum_{i=1}^N \left[ \pi_i^2 + m^2 \phi_i^2 \right] + \frac{\lambda}{4 N} \sum_{i,j=1}^N \phi_i^2 \phi_j^2 + \sum_{i=1}^N f_i(t) \phi_i, \\
\pd_t \rho = -i \left[ H, \rho \right] + \Gamma \sum_{i=1}^N \left( a_i \rho a_i^\dag - \frac{1}{2} a_i^\dag a_i \rho - \frac{1}{2} \rho a_i^\dag a_i \right),
\end{gathered} \eeq
where $\rho$ denotes the density matrix of the system and $\Gamma$ determines the dissipation rate. Since this model resembles the large-$N$ generalization of the Duffing oscillator, we expect it to exhibit quantum and classical chaos for some parameters of the model and driving forces. We will consider this model elsewhere.

Finally, note that the model~\eqref{eq:S} is very similar to some spatially reduced string and gauge models~\cite{Gur-Ari, Buividovich, Asano, Berkowitz, Kawahara, Matinyan, Chirikov, Savvidy:1984, Biro, Kunihiro, Baskan, Savvidy:2022}. In particular, the $N=3$ variant of~\eqref{eq:S} is nothing but a spatially reduced $SU(2)$ Yang-Mills~\cite{Matinyan}, which is known to possess a nonzero classical Lyapunov exponent $\kappa \sim \sqrt[4]{E}$, see~\cite{Chirikov, Savvidy:1984, Kunihiro, Savvidy:2022}. Besides, the model~\eqref{eq:S} is very similar to the model of a strongly coupled phonon fluid~\cite{Tulipman-1, Tulipman-2}, where the $O(N)$ symmetry is broken in a slightly different way. Due to these reasons, we expect our analysis to provide useful insights into the physics of these complex models.

\section*{Acknowledgments}

We thank Emil Akhmedov, Alexander Gorsky, Andrei Semenov, and Rustem Sharipov for the fruitful discussions. The work of DAT was supported by the grant from the Foundation for the Advancement of Theoretical Physics and Mathematics ``BASIS''. The work of NK was supported by the RFBR grant No.20-02-00297
and by the Foundation for the Advancement of Theoretical Physics and Mathematics ``BASIS''.

\appendix

\section{Schwinger-Keldysh technique on a two-fold contour}
\label{sec:technique}

\subsection{Basic rules}
\label{sec:SK-basics}

In the main body of the paper, we need to calculate the following regularized correlation functions:
\beq \label{eq:basic-correlator}
\tilde{C}(t_1,t_2,t_3,t_4) = \tr \left[ U(\infty, t_0) \rho^{1/2} U^\dag(\infty, t_0) \phi(t_1) \phi(t_2) U(\infty, t_0) \rho^{1/2} U^\dag(\infty, t_0) \phi(t_3) \phi(t_4) \right], \eeq
where $\rho$ is the density matrix at the moment $t_0$ and $U(t,t_0) = \mT \exp\left[ -i \int_{t_0}^t H_\text{int}(t') dt' \right]$ is the evolution operator in the interaction picture. In this section, we consider the $O(N)$-symmetric version of the model~\eqref{eq:S} and suppress the group indices of $\phi$ for brevity. The generalization to the full nonsymmetric model is straightforward.

Expanding the evolution operators in the powers of $\lambda$ and using Wick's theorem\footnote{We remind that this theorem works only if we assume that the initial Hamiltonian is Gaussian. Otherwise, more complex correlations (e.g., double correlations that appear when $\langle a_1^\dag a_2^\dag a_1 a_2 \rangle \neq \langle a_1^\dag a_1 \rangle \langle a_2^\dag a_2 \rangle$) should be taken into account, and the diagrammatic technique should be augmented with corresponding correlation blocks~\cite{Arseev, Hall, Kukharenko, Radovskaya}.}, we straightforwardly rewrite these complex correlation functions as products of two-point correlators time-ordered along the two-fold Keldysh contour $\mathcal{C}$ (Fig.~\ref{fig:contour}). The main motivation to introduce this contour is an unambiguous specification of the operator ordering in an arbitrary four-point correlator~\eqref{eq:basic-correlator}, which is achieved by assigning the operators to the appropriate branches of the contour. So, the correlator can be conveniently rewritten as follows:
\begin{figure}[t]
    \center{\includegraphics[scale=0.37]{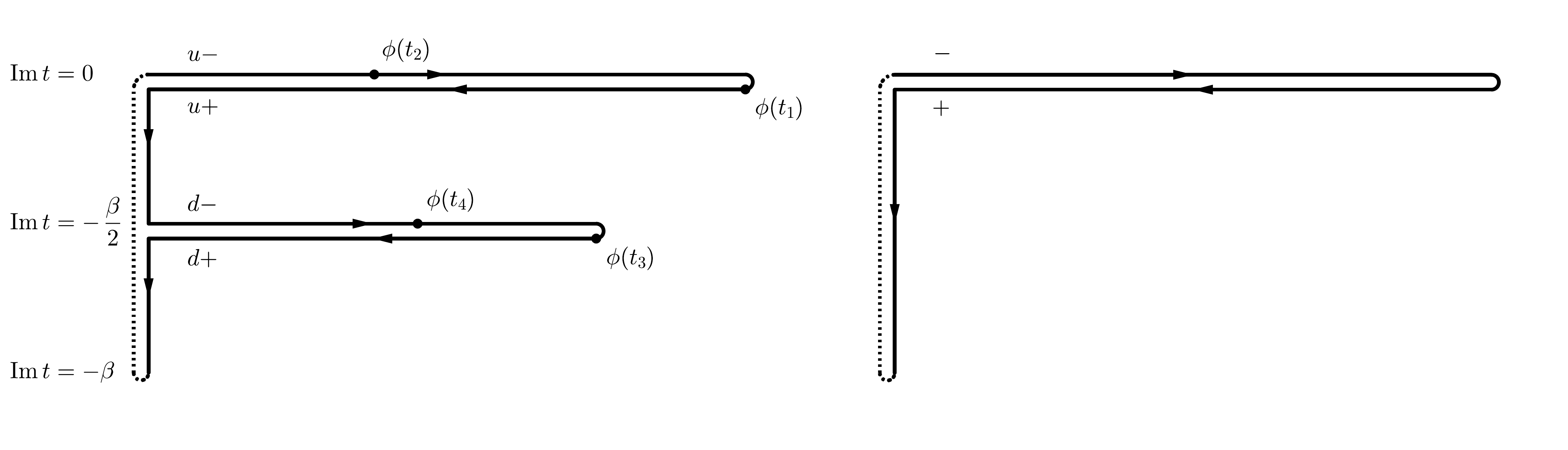}}
    \caption{Two-fold (left) and conventional (right) Keldysh contour $\mathcal{C}$ for a thermal system. Horizontal lines denote the evolution in the real time, vertical lines denote the evolution in the imaginary time, and dotted lines denote the identification of points $t = t_0$ and $t = t_0 - i \beta$. }
    \label{fig:contour}
\end{figure}
\beq \label{eq:pm-correlator} \begin{aligned}
\tilde{C}(t_1,t_2,t_3,t_4) &= \tr \left[ \mT_\mathcal{C} \, \rho \, \phi_{u+}(t_1) \phi_{u-}(t_2) \phi_{d+}(t_3) \phi_{d-}(t_4) \exp\left( - i \int_\mathcal{C} H_\text{int}(t') dt'\right) \right] \\ &\equiv \left\langle \phi_{u+}(t_1) \phi_{u-}(t_2) \phi_{d+}(t_3) \phi_{d-}(t_4) \right\rangle. \end{aligned} \eeq
Essentially, the ``minus'' (``plus'') branches are generated by the forward (backward) time evolution, i.e., by the operators $U^{(\dag)}(\infty, t_0)$. Therefore, in this notation, we obtain four different interaction vertices (we remind that we consider the $O(N)$-symmetric model with a quartic interaction):
\beq \pm i \frac{\lambda}{4 N} \int_{t_0}^\infty \phi_{u\pm}^4(t') dt', \quad \pm i \frac{\lambda}{4 N} \int_{t_0}^\infty \phi_{d\pm}^4(t') dt', \eeq
and sixteen propagators, $G_{\alpha,\beta}(t,t') \equiv -i \langle \phi_\alpha(t) \phi_\beta(t') \rangle$, which are conveniently collected in a $4 \times 4$ matrix:
\beq \label{eq:pm-propagators}
\mathbb{G}(t,t') = \bem G_{u-,u-}(t,t') & G_{u-,u+}(t,t') & G_{u-,d-}(t,t') & G_{u-,d+}(t,t') \\ G_{u+,u-}(t,t') & G_{u+,u+}(t,t') & G_{u+,d-}(t,t') & G_{u+,d+}(t,t') \\ G_{d-,u-}(t,t') & G_{d-,u+}(t,t') & G_{d-,d-}(t,t') & G_{d-,d+}(t,t') \\ G_{d-,u-}(t,t') & G_{d-,u+}(t,t') & G_{d-,d-}(t,t') & G_{d-,d+}(t,t') \eem. \eeq
However, this cumbersome notation can be significantly simplified because almost all propagators are linearly dependent. Indeed, keeping in mind that empty forward and backward branches cancel each other (simply put, $U^\dag(\infty, t) U(\infty, t) = 1$) and that contour $\mathcal{C}$ is invariant under the cyclic permutation of the upper and bottom folds, one can easily infer the following relations (arguments of propagators are suppressed for brevity):
\beq \label{eq:correlator-relations}
\begin{gathered}
G_{u-,u-} = G_{d-,d-}, \quad G_{u-,u+} = G_{d-,d+}, \quad G_{u+,u-} = G_{d+,d-}, \quad G_{u+,u+} = G_{d+,d+}, \\
G_{u-,d-} = G_{u-,d+} = G_{u+,d-} = G_{u+,d+} = G_{d-,u-} = G_{d-,u+} = G_{d+,u-} = G_{d+,u+}.
\end{gathered} \eeq
Moreover, the propagators on the same fold are also dependent:
\beq G_{u-,u+} + G_{u+,u+} = G_{u-,u+} + G_{u+,u-}, \quad  G_{d-,d+} + G_{d+,d+} = G_{d-,d+} + G_{d+,d-}. \eeq
Hence, it is convenient to rotate from the ``$\pm$'' components to the so-called ``classical'' and ``quantum'' ones:
\beq \label{eq:cq}
\bem \phi_{uc} \\ \phi_{uq} \eem = R \bem \phi_{u-} \\ \phi_{u+} \eem, \quad \bem \phi_{dc} \\ \phi_{dq} \eem = R \bem \phi_{d-} \\ \phi_{d+} \eem, \quad R = \bem \frac{1}{2} & \frac{1}{2} \\ 1 & -1 \eem, \eeq
and introduce four linearly independent propagators instead of sixteen dependent ones:
\beq \label{eq:propagators-app}
\begin{aligned}
G^R(t,t') &= -i \langle \phi_{uc}(t) \phi_{uq}(t') \rangle = -i \langle \phi_{dc}(t) \phi_{dq}(t') \rangle, \\
G^A(t,t') &= -i \langle \phi_{uq}(t) \phi_{uc}(t') \rangle = -i \langle \phi_{dq}(t) \phi_{dc}(t') \rangle, \\
G^K(t,t') &= -i \langle \phi_{uc}(t) \phi_{uc}(t') \rangle = -i \langle \phi_{dc}(t) \phi_{dc}(t') \rangle, \\
G^W(t,t') &= -i \langle \phi_{uc}(t) \phi_{dc}(t') \rangle = -i \langle \phi_{dc}(t) \phi_{uc}(t') \rangle.
\end{aligned} \eeq
We refer to these correlators as retarded, advanced, Keldysh, and Wightman propagators, respectively. Note that the quantum-quantum correlators between any folds and classical-quantum correlators between different folds are zero due to identities~\eqref{eq:correlator-relations}, e.g.:
\beq -i \langle \phi_{uc}(t) \phi_{dq}(t') \rangle = G_{u-,d-}(t,t') + G_{u+,d-}(t,t') - G_{u-,d+}(t,t') - G_{u+,d+}(t,t') = 0. \eeq
This significantly restricts the number of diagrams in the perturbative expansion of~\eqref{eq:basic-correlator}.

\begin{figure}[t]
    \center{\includegraphics[scale=0.4]{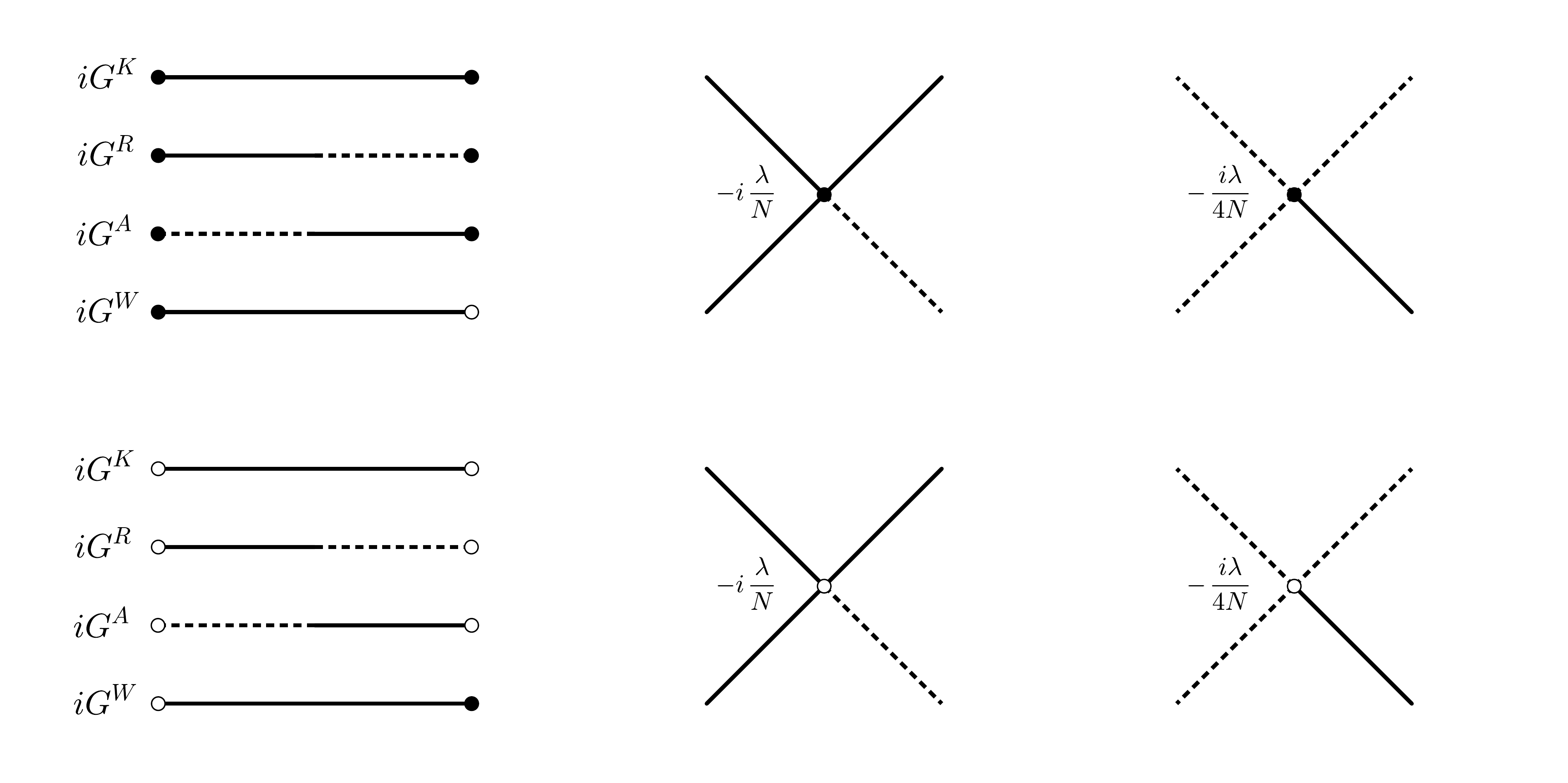}}
    \caption{Propagators and vertices in the augmented Schwinger-Keldysh diagrammatic technique of the $O(N)$-symmetric model on a two-fold contour. The solid and dashed lines correspond to the classical and quantum components; the black and white dots correspond to the fields on the upper and lower folds; the group indices are suppressed for brevity. Note that the full nonsymmetric model~\eqref{eq:S} contains twice more interaction vertices (cf. Fig.~\ref{fig:vertices}).}
    \label{fig:technique}
\end{figure}

Substituting the rotated fields into~\eqref{eq:pm-correlator} and regrouping the integrals in the exponential function, we also obtain the vertices in the ``cq'' notation:
\beq \label{eq:vertices}
\begin{aligned}
&-i \frac{\lambda}{N} \int_{t_0}^\infty \phi_{uc}^3(t') \phi_{uq}(t') dt', &\quad &- i \frac{\lambda}{4 N} \int_{t_0}^\infty \phi_{uc}(t') \phi_{uq}^3(t') dt', \\ &-i \frac{\lambda}{N} \int_{t_0}^\infty \phi_{dc}^3(t') \phi_{dq}(t') dt', &\quad &-i \frac{\lambda}{4 N} \int_{t_0}^\infty \phi_{dc}(t') \phi_{dq}^3(t') dt'.
\end{aligned} \eeq
Finally, let us write down the expectation value of the squared commutator~\eqref{eq:C} in the ``cq'' notation (we assume that $t_1 > t_3$ and $t_2 > t_4$):
\beq \begin{aligned}
C(t_1, t_2; t_3, t_4) &= - \tr \left\{ U(\infty, t_0) \rho^{1/2} U^\dag(\infty, t_0) \left[ \phi(t_1), \phi(t_3) \right] U(\infty, t_0) \rho^{1/2} U^\dag(\infty, t_0) \left[ \phi(t_2), \phi(t_4) \right] \right\} \\
&= - \left\langle \phi_{u+}(t_1) \phi_{u-}(t_3) \phi_{d+}(t_2) \phi_{d-}(t_4) \right\rangle - \left\langle \phi_{u-}(t_1) \phi_{u+}(t_3) \phi_{d-}(t_2) \phi_{d+}(t_4) \right\rangle \\
&\quad+ \left\langle \phi_{u+}(t_1) \phi_{u-}(t_3) \phi_{d-}(t_2) \phi_{d+}(t_4) \right\rangle + \left\langle \phi_{u-}(t_1) \phi_{u+}(t_3) \phi_{d+}(t_2) \phi_{d-}(t_4) \right\rangle \\
&= -\left\langle \phi_{uc}(t_1) \phi_{uq}(t_3) \phi_{dc}(t_2) \phi_{dq}(t_4) \right\rangle.
\end{aligned} \eeq
Here, we use the identity $U^\dag(\infty,t) U(\infty,t)$ to cut the folds at times $t_1$ and $t_2$ and move the operators $\phi(t_1)$ and $\phi(t_2)$ between the ``$-$'' and ``$+$'' branches of the corresponding folds. In the zeroth order, this correlator is just a product of two bare retarded propagators (all other contractions are zero):
\beq \begin{aligned}
C(t_1, t_2; t_3, t_4) &= - \left\langle \phi_{uc}(t_1) \phi_{uq}(t_3) \phi_{dc}(t_2) \phi_{dq}(t_4) \right\rangle_{\lambda = 0} \\ &= - \left\langle \phi_{uc}(t_1) \phi_{uq}(t_3) \right\rangle_{\lambda = 0} \left\langle \phi_{dc}(t_2)  \phi_{dq}(t_4) \right\rangle_{\lambda = 0} = G_0^R(t_1, t_3) G_0^R(t_2, t_4).
\end{aligned} \eeq
At nonzero $\lambda$, corrections to this expression are described by the augmented Schwinger-Keldysh technique on the two-fold contour $\mathcal{C}$. The rules of this technique follow from Eqs.~\eqref{eq:propagators-app} and~\eqref{eq:vertices}, see Fig.~\ref{fig:technique}. More details on the derivation and applications of the generalized Schwinger-Keldysh technique can be found in~\cite{Aleiner,Haehl}.

\subsection{Bare propagators}
\label{sec:SK-propagators}

Now let us determine the tree-level propagators in the quantized model~\eqref{eq:S}. First of all, we remind that the quantized field is decomposed as follows:
\beq \phi_i(t) = a_i f(t) + a_i^\dag f^*(t), \eeq
Here, the creation and annihilation operators satisfy the commutation relations $\big[ a_i, a_j^\dag \big] = \delta_{ij}$, and the mode function $f(t) = \frac{1}{\sqrt{2 m}} e^{-i m t}$ solves the classical equation of motion and ensures the canonical commutation relations, $\big[ \phi_j(t), \pi_k(t) \big] = i \delta_{jk} $ (note that the canonical momentum $\pi_i = \dot{\phi}_i$). The Hamiltonian at the initial moment $t_0$, where interactions are turned off, also has a simple form:
\beq H(t_0) = H_\text{free}(t_0) = m \left( a_i^\dag a_i + \frac{N}{2} \right). \eeq
Hence, the initial density matrix, which describes a thermal system at the inverse temperature $\beta$, is as follows:
\beq \rho = \frac{1}{Z} e^{-\beta H(t_0)}, \quad \text{where} \quad Z = \tr\left[ e^{-\beta H(t_0)} \right] = \frac{e^{-\beta m N/2}}{1 - e^{-\beta m}}. \eeq
Using these relations, we readily calculate the following traces:
\beq \begin{gathered}
    \tr\left[ \rho \, a_i^\dag a_j \right] = 
    \frac{1}{e^{\beta m} - 1} \delta_{ij}, \qquad
    \tr\left[ \rho \, a_i^\dag a_j \right] = \frac{e^{\beta m}}{e^{\beta m} - 1} \delta_{ij}, \\
    \tr\left[ \rho^{1/2} a_i^\dag \rho^{1/2} a_j \right] = \frac{e^{\beta m / 2}}{e^{\beta m} - 1} \delta_{ij}.
\end{gathered} \eeq
Finally, rewriting the ``cq'' propagators~\eqref{eq:propagators-app} through the ``$\pm$'' propagators~\eqref{eq:pm-propagators}, keeping in mind the ordering of operators in the ``$\pm$'' propagators and substituting the explicit form of the mode function $f(t)$, we obtain the bare two-point correlation functions~\eqref{eq:bare-propagators}.

We emphasize that the tree-level retarded and advanced propagators do not depend on the inverse temperature. The dependence disappears because these propagators are expressed through the commutator of two fields, which is a c-number. Please also note that in the limit $\beta m \gg 1$, i.e., at very low temperatures, the Wightman propagator is exponentially suppressed. In other words, in this limit, correlations between the different folds are negligible.

\subsection{Relation to Matsubara propagators}
\label{sec:Matsubara}

In general, calculations in the Schwinger-Keldysh technique are lengthy and tedious due to the large number of possible diagrams (e.g., the two-loop correction to the Keldysh propagator in the $\lambda \phi^4$ model, Fig.~\ref{fig:technique}, already involves seven different diagrams). However, these calculations are redundant if initial quantum state is thermal. In this case, the exact propagators~\eqref{eq:propagators-app} are derived from the Matsubara propagators using the analytic continuation procedure. In this subsection, we briefly review the Matsubara technique and reproduce propagators~\eqref{eq:bare-propagators}.

Essentially, the Matsubara technique describes the evolution of a system in the imaginary time, $\tau = i t$:
\beq \left\langle \mT_\tau \phi_i(\tau_1) \cdots \phi_j(\tau_2) \right\rangle_\tau \equiv \frac{\tr\left[ \mT_\tau \phi_i(\tau_1) \cdots \phi_j(\tau_2) U(\beta) e^{-\beta H_\text{free}} \right]}{\tr\left[U(\beta) e^{-\beta H_\text{free}} \right]}. \eeq
Here, $\mT_\tau$ denotes the time-ordering in the imaginary time, $\beta$ is the inverse temperature, $H_\text{free}$ is the unperturbed Hamiltonian and $U(\tau) = \mT_\tau \exp\left[- \int_0^\tau H_\text{int}(\tau') d\tau'\right]$ is the evolution operator in the interaction picture. Similarly to the real-time technique described in subsection~\ref{sec:SK-basics}, we can rewrite many-point correlation functions as the products of two-point correlators:
\beq G_{ij}(\tau) = - \left\langle \mT_\tau \phi_i(\tau) \phi_j(0) \right\rangle_\tau. \eeq
However, note that the evolution in the imaginary time is restricted to the interval $0 < \tau < \beta$; moreover, the states of the system at the moments $\tau = 0$ and $\tau = \beta$ coincide. This implies the periodic boundary conditions on the correlation functions:
\beq G_{ij}(\tau + \beta) = G_{ij}(\tau). \eeq
These conditions mean that the frequency in the Fourier-transformed propagator takes only discrete values:
\beq G_{ij}(\tau) = \frac{1}{\beta} \sum_{\omega_n} G_{ij}(i \omega_n) e^{-i \omega_n \tau}, \quad \text{where} \quad \omega_n = \frac{2 \pi n}{\beta}, \quad n \in \mathbb{Z}. \eeq
Hence, the frequency-space formulation of the Matsubara technique contains the sums over the loop frequencies ($\frac{1}{\beta} \sum_{\omega_n}$). Otherwise, this diagrammatic technique is very similar to the Feynman one.

Furthermore, the imaginary-time and real-time, Eq.~\eqref{eq:propagators-app}, propagators are related via the analytic continuation of the frequency~\cite{Abrikosov, Lifshitz, Rammer}:
\beq \label{eq:FDT}
\begin{gathered}
G^R(\omega) = -G(i \omega_n \to \omega + i 0), \quad G^A(\omega) = -G(i \omega_n \to \omega - i 0), \\
G^K(\omega) = \frac{1}{2} \coth \frac{\beta \omega}{2} \left[ G^R(\omega) - G^A(\omega) \right], \\
G^W(\omega) = \frac{1}{2} \csch \frac{\beta \omega}{2} \left[ G^R(\omega) - G^A(\omega) \right].
\end{gathered} \eeq
Here, we denote $\csch(x) = 1/\sinh(x)$ for brevity. The relation between the Keldysh, retarded, and advanced propagators is usually referred to as the fluctuation-dissipation theorem. The derivation of the relation for the Wightman propagator on a two-fold contour can be found, e.g., in appendix~C of~\cite{Swingle}.

Let us derive the tree-level propagators~\eqref{eq:bare-propagators} from the Euclidean version of the model~\eqref{eq:S}:
\beq \label{eq:S-E}
S_E = \int_0^\beta d\tau' \left[ \frac{1}{2} \dot{\phi}_i^2 + \frac{m^2}{2} \phi_i^2 + \frac{\lambda}{4 N} \left( \phi_i^2 \phi_j^2 - \phi_i^4 \right) \right]. \eeq
It is straightforward to see that in this model, the bare propagator has the following form:
\beq G_{0;ij}(i \omega_n) = G_0(i \omega_n) \delta_{ij}, \quad G_0(i \omega_n) = \frac{1}{\omega_n^2 + m^2}. \eeq
Hence, the bare real-time propagators are:
\beq \begin{aligned}
i G_0^R(t) &= \int_{-\infty}^\infty \frac{i}{(\omega + i 0)^2 - m^2} e^{-i \omega t} \frac{d\omega}{2 \pi} = - i \theta(t) \frac{\sin \left( m t \right)}{m}, \\
i G_0^A(t) &= \int_{-\infty}^\infty \frac{i}{(\omega - i 0)^2 - m^2} e^{-i \omega t} \frac{d\omega}{2 \pi} = i \theta(-t) \frac{\sin \left( m t \right)}{m}, \\
i G_0^K(t) &= \int_{-\infty}^\infty \frac{1}{2} \coth \left( \frac{\beta \omega}{2} \right) \frac{\pi}{m} \left[ \delta(\omega - m) - \delta(\omega + m) \right] e^{-i \omega t} \frac{d\omega}{2 \pi} = \frac{1}{2} \coth \frac{\beta m}{2} \, \frac{\cos \left( m t \right)}{m}, \\
i G_0^W(t) &= \int_{-\infty}^\infty \frac{1}{2} \csch\left( \frac{\beta \omega}{2} \right) \frac{\pi}{m} \left[ \delta(\omega - m) - \delta(\omega + m) \right] e^{-i \omega t} \frac{d\omega}{2 \pi} = \frac{e^{\beta m/2}}{e^{\beta m}- 1} \frac{\cos \left( m t \right)}{m}.
\end{aligned} \eeq
This reproduces identities~\eqref{eq:bare-propagators} with $t = t_1 - t_2$.

Finally, we emphasize that similar relations also hold for the exact imaginary-time and real-time two-point correlation functions. Nevertheless, this correspondence cannot be straightforwardly extended to OTOCs, which depend on times from different folds of the Keldysh contour, i.e., describe a mixed evolution in both imaginary and real times.

\section{The right way to introduce Langrange fields}
\label{sec:langrange}

A conventional approach to solve the $O(N)$ model is to do the Hubbard-Stratonovich transformation and introduce the Lagrange field $\sigma(t)$, e.g., see~\cite{Moshe, Polyakov:book, Coleman, Schubring}. The starting point of this approach is the Euclidean path integral representation of the partition function:
\beq \label{eq:Z-Lagrange}
\begin{aligned}
Z &= \int \mD \phi_i \exp\left[ - \int_0^\beta d\tau \left( \frac{1}{2} \dot{\phi}_i^2 + \frac{m^2}{2} \phi_i^2 + \frac{\lambda}{4 N} \phi_i^2 \phi_j^2 \right) \right] \\ &\to \int \mD \phi_i \mD \sigma \exp\left[ - \int_0^\beta d\tau \left( \frac{1}{2} \dot{\phi}_i^2 + \frac{m^2}{2} \phi_i^2 - \frac{1}{\sqrt{N}} \sigma \phi_i^2 - \frac{1}{\lambda} \sigma^2 \right) \right].
\end{aligned} \eeq
Note that integrals in the first and second lines differ by a constant factor, which is insignificant for calculating correlation functions. So, using this representation, we find the tree-level Matsubara propagator of the Lagrange field:
\beq D_0(i \omega_n) = -\frac{\lambda}{2}. \eeq
Taking in mind the diagrammatic rules that follow from~\eqref{eq:Z-Lagrange}, we also straightforwardly derive the Dyson-Schwinger equations:
\beq \begin{gathered}
G^{-1}(i \omega_n) = G_0^{-1}(i \omega_n) - \frac{2}{\beta} \sum_{\omega_k} D_0(0) G(i \omega_k) + \mO\left( \frac{1}{N} \right), \\
D^{-1}(i \omega_n) = D_0^{-1}(i \omega_n) - \frac{2}{\beta} \sum_{\omega_k} G(i \omega_k) G(i \omega_{n+k}) + \mO\left( \frac{1}{N} \right),
\end{gathered} \eeq
and calculate the leading order resummed Matsubara propagators of the fields $\phi_i$ and $\sigma$,
\beq \label{eq:propagators-Lagrange}
\begin{gathered}
G(i \omega_n) = \frac{1}{\omega_n^2 + \tilde{m}^2}, \\
D(i \omega_n) = -\frac{\lambda}{2} + \frac{\lambda \left(\tilde{m}^2 - m^2\right)}{\omega_n^2 + 6 \tilde{m}^2 - 2 m^2} \approx -\frac{\lambda}{2} + \frac{\lambda \tilde{m}^2}{\omega_n^2 + 6 \tilde{m}^2}.
\end{gathered} \eeq
Here, we use the group structure of $\phi \phi$ propagators, $G_{ij}(i \omega_n) = G(i \omega_n) \delta_{ij}$, and neglect the subleading, $\mO(1/N)$, corrections to the propagators. The resummed mass $\tilde{m}$ is again determined by Eq.~\eqref{eq:thermal-m}. The approximate identity in~\eqref{eq:propagators-Lagrange} is established in the high-temperature limit, $\beta m \ll 1$ and $\beta m \ll \lambda/m^3$.

Since we need the real-time propagators to estimate the resummed four-point correlator~\eqref{eq:C-averaged}, we analytically continue Matsubara propagators~\eqref{eq:propagators-Lagrange} to real frequencies and do a Fourier transform:
\beq \begin{gathered}
i G^R(t) = - i \theta(t) \frac{\sin\left( \tilde{m} t \right)}{\tilde{m}}, \\
i D^R(t) = i \frac{\lambda}{2} \delta(t) - i \frac{\lambda}{2} \left( \mu - \frac{4}{\mu} \right) \tilde{m} \theta(t) \sin\left( \mu \tilde{m} t \right),
\end{gathered} \eeq
where $\mu = \sqrt{6 - 2 m^2/\tilde{m}^2}$. These propagators coincide with retarded propagators~\eqref{eq:tadpoles} and~\eqref{eq:bubble-chain-result} calculated in the initial model~\eqref{eq:S} (note that in the leading order in $1/N$, this model coincides with the $O(N)$-symmetric one). Furthermore, using the fluctuation-dissipation theorem and its analog for Wightman propagators, Eq.~\eqref{eq:FDT}, we restore the correct $\mO(1)$ resummed Keldysh and Wightman $\phi \phi$ propagators:
\beq \begin{gathered}
i G^K(t) = \frac{1}{2} \coth \frac{\beta \tilde{m}}{2} \frac{\cos\left(\tilde{m} t\right)}{\tilde{m}}, \\
i G^W(t) = \frac{e^{\beta \tilde{m}/2}}{e^{\beta \tilde{m}} - 1} \frac{\cos\left(\tilde{m} t\right)}{\tilde{m}}.
\end{gathered} \eeq
So it is tempting to assume that the Lagrange field thermalizes and apply the fluctuation-dissipation theorem to $\sigma \sigma$ propagators as well~\cite{Swingle}:
\beq \label{eq:false-propagators}
\begin{gathered}
i D^K(t) = \frac{\lambda}{2} \frac{1}{2} \coth \frac{\beta \mu \tilde{m}}{2} \left( \mu - \frac{4}{\mu} \right) \tilde{m} \cos\left( \mu \tilde{m} t \right), \\
i D^W(t) = \frac{\lambda}{2} \frac{e^{\beta \mu \tilde{m}/2}}{e^{\beta \mu \tilde{m}} - 1} \left( \mu - \frac{4}{\mu} \right) \tilde{m} \cos\left( \mu \tilde{m} t \right).
\end{gathered} \eeq
However, this approach is misleading because the initial state of the Lagrange field is \textit{not thermal}; moreover, it \textit{does not thermalize} due to the properties of the model~\eqref{eq:S} (see Appendix~\ref{sec:subleading-self-energy}). To restore the correct initial state of the Lagrange field, we consider the real-time partition function of the $O(N)$ symmetric model~\cite{Berges, Leonidov-1, Leonidov-2, Radovskaya}:
\beq Z = \int \mD \varphi_i \mD \pi_i \, W\left[ \varphi_i, \pi_i \right] \int_{i.c.} \mD \phi_{i,c}(t) \mD \phi_{i,q}(t) e^{i S_K\left[ \phi_{i,c}(t), \phi_{i,q}(t) \right]}, \eeq
where $S_K$ denotes the Keldysh action after the Keldysh rotation:
\beq S_K\left[ \phi_{i,c}(t), \phi_{i,q}(t) \right] = -\int_{t_0}^\infty \left[ \phi_{i,q} \left( \pd_t^2 + m^2 \right) \phi_{i,c} + \frac{\lambda}{N} \phi_{i,c} \phi_{i,c} \phi_{j,c} \phi_{j,q} + \frac{\lambda}{4 N} \phi_{i,c} \phi_{i,q} \phi_{j,q} \phi_{j,q} \right], \eeq
$W\left[ \varphi_i, \pi_i \right]$ denotes the Wigner function, which is related to the initial value of the density matrix:
\beq \label{eq:Wigner}
W\left[ \varphi_i, \pi_i \right] = \int \mD \alpha_i \, e^{i \alpha_i \pi_i} \Big\langle \varphi_i + \frac{1}{2} \alpha_i \Big| \rho(t_0) \Big| \varphi_i - \frac{1}{2} \alpha_i \Big\rangle, \eeq
and the integral with ``$i.c.$'' means the initial values for the classical fields, $\phi_{i,c}(t_0) = \varphi_i$, $\dot{\phi}_{i,c}(t_0) = \pi_i$, whereas the initial values for the quantum fields are not fixed. Similarly to the Eculidean case~\eqref{eq:Z-Lagrange}, we introduce the two-component Lagrange field to get rid of the quartic interaction term:
\beq Z = \int \mD \varphi_i \mD \pi_i \mD \sigma_0 \, W\left[ \varphi_i, \pi_i \right] \int_{i.c.} \mD \phi_{i,c}(t) \mD \phi_{i,q}(t) \mD \sigma_c(t) \mD \sigma_q(t) e^{i \tilde{S}_K\left[ \phi_{i,c}(t), \phi_{i,q}(t), \sigma_c(t), \sigma_q(t) \right]}, \eeq
where the transformed Keldysh action is as follows:
\beq \tilde{S}_K = -\int_{t_0}^\infty \left[ \phi_{i,q} \left( \pd_t^2 + m^2 \right) \phi_{i,c} - \frac{2}{\lambda} \sigma_c \sigma_q - \frac{1}{\sqrt{N}} \sigma_c \phi_{i,c} \phi_{i,q} - \frac{1}{\sqrt{N}} \sigma_q \phi_{i,c} \phi_{i,c} - \frac{1}{4 \sqrt{N}} \sigma_q \phi_{i,q} \phi_{i,q} \right], \eeq
and the classical Lagrange field satisfies the initial condition $\sigma_c(t_0) = \sigma_0$. Note that the canonical momentum of this field is ill-defined. Moreover, the full Wigner function in the transformed theory, which incorporates both $\phi_i$ and $\sigma$, coincides with the Wigner function~\eqref{eq:Wigner} before the transformation. Hence, the Lagrange field cannot possess any initial quantum correlations:
\beq W\left[ \phi_i, \pi_i, \sigma_0 \right] = \int \mD \alpha_i \, e^{i \alpha_i \pi_i} \bigg\langle \bem \varphi_i + \frac{1}{2} \alpha_i \\ \sigma_0 \eem^\mathrm{T} \bigg| \bem \rho(t_0) & 0 \\ 0 & 0 \eem \bigg| \bem \varphi_i - \frac{1}{2} \alpha_i \\ \sigma_0 \eem \bigg\rangle = W\left[ \phi_i, \pi_i \right]. \eeq
The absence of initial correlations means that the tree-level Keldysh and Wightman\footnote{For simplicity, in this appendix, we restrict ourselves to the standard one-fold Keldysh contour (right picture on Fig.~\ref{fig:contour}). However, our reasoning is straightforwardly extended to the case of a two-fold contour (left picture on Fig.~\ref{fig:contour}). Such an extension additionally establishes the absence of initial correlations between different folds.} propagators of the Lagrange field are zero:
\beq D_0^K(t) = 0, \qquad D_0^W(t) = 0. \eeq
At the same time, the tree-level retarded and advanced propagators do not depend on the initial state and can be safely derived from the Euclidean model.

Finally, taking into account all these observations, we restore the correct $\mO(1)$ resummed Keldysh and Wightman propagators in the $O(N)$-symmetric model:
\beq i D^K(t) = i D^W(t) = \frac{1}{2 \mu^4} \frac{\lambda^2}{\tilde{m}^2} \left( \csch \frac{\beta \tilde{m}}{2} \right)^2. \eeq
Obviously, these propagators do not conicide with the naive ones~\eqref{eq:false-propagators}.

\section{The subleading correction to the self-energy}
\label{sec:subleading-self-energy}

\begin{figure}[t]
    \center{\includegraphics[scale=0.37]{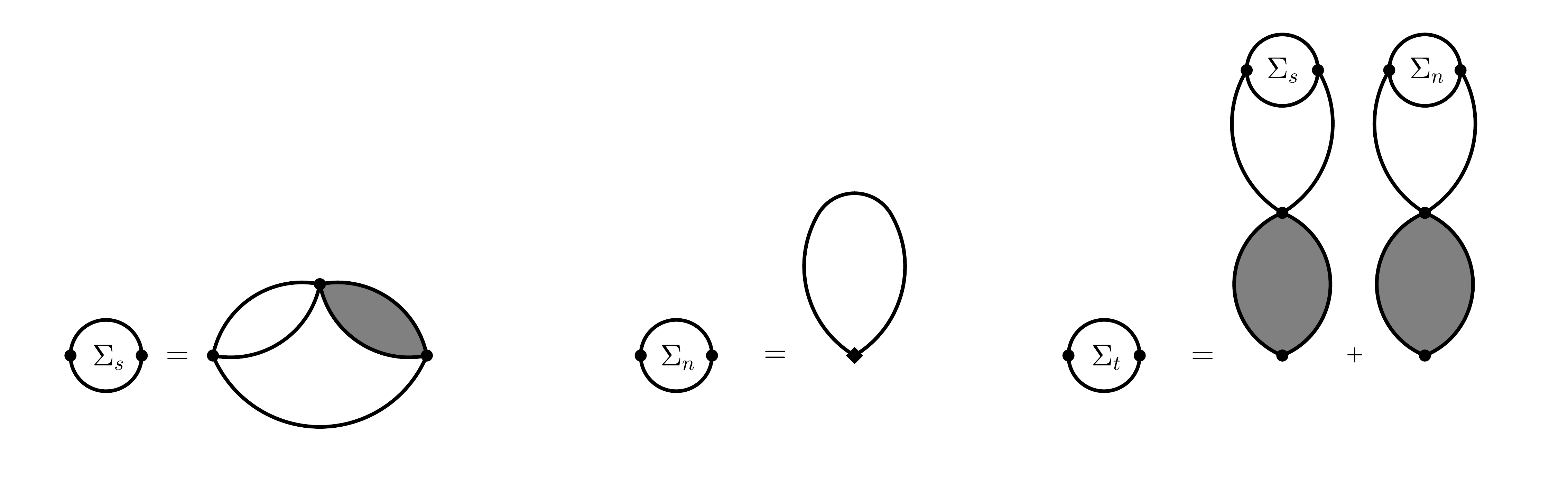}}
    \caption{Contributions to the self-energy in the $1/N$ order in Matsubara technique. The lines correspond to the $\mO(1)$ resummed tadpoles diagrams, the shaded loops correspond to the $\mO(1)$ resummed chain of bubble diagrams~\eqref{eq:bubble-chain-matsubara}.}
    \label{fig:self-energy}
\end{figure}

In this appendix, we calculate the $1/N$ correction to the self-energy of the $\phi \phi$ propagator and show that it does not contain imaginary contributions. As was explained in Appendix~\ref{sec:Matsubara}, we can safely use the Matsubara technique for this purpose. So, in these notations, the Dyson-Schwinger equation on the $\mO(1/N)$ resummed propagator $\tilde{G}(i \omega_n)$ has the following form:
\beq \tilde{G}(i \omega_n) = G(i \omega_n) + G(i \omega_n) \Sigma(i \omega_n) \tilde{G}(i \omega_n), \eeq
where $G(i \omega_n)$ denotes the $\mO(1)$ resummed propagator and $\Sigma(i \omega_n)$ denotes the
$1/N$ contribution to the self-energy. For convenience, we separate the terms that correspond to different diagrams (see Fig.~\ref{fig:self-energy}):
\beq \Sigma(i \omega_n) = \Sigma_s(i \omega_n) + \Sigma_n(i \omega_n) + \Sigma_t(i \omega_n), \eeq
Here, $\Sigma_s(i \omega_n)$ describes the nonlocal ``sunset'' contribution:
\beq \label{eq:self-energy-sun}
\begin{aligned}
\Sigma_s(i \omega_n) &= \frac{2}{N} \frac{\lambda^2}{\beta^2} \sum_{\omega_k, \omega_l} \frac{1}{(\omega_n + \omega_k)^2 + \tilde{m}^2} \frac{1}{(\omega_k + \omega_l)^2 + \tilde{m}^2} \frac{1}{\omega_l^2 + \tilde{m}^2} B(i \omega_k) \\ &\approx \frac{1}{N} \frac{2}{3} \frac{\left( 7 \omega_n^2 + 25 \tilde{m}^2 \right) \tilde{m}^4}{\omega_n^4 + 14 \omega_n^2 \tilde{m}^2 + 25 \tilde{m}^4},
\end{aligned} \eeq
$\Sigma_n(i \omega_n)$ corresponds to the tadpole contribution with an $O(N)$-nonsymmetric vertex:
\beq \label{eq:self-energy-nsymm} \Sigma_n(i \omega_n) = \frac{3}{N} \frac{\lambda}{\beta} \sum_{\omega_k} \frac{1}{\omega_k^2 + \tilde{m}^2} = \frac{3}{N} \frac{\lambda}{2 \tilde{m}} \coth \frac{\beta \tilde{m}}{2} \approx \frac{1}{N} 3 \tilde{m}^2, \eeq
and $\Sigma_t(i \omega_n)$ incorporates the $1/N$ corrections to the tadpole diagrams:
\beq \label{eq:self-energy-tadpole} \Sigma_t(i \omega_n) = -\frac{\lambda}{\beta} \sum_{\omega_k} \frac{B(i \omega_k) \Sigma_s(i \omega_k)}{\left( \omega_k^2 + \tilde{m}^2 \right)^2 } - \frac{\lambda}{\beta} \sum_{\omega_k} \frac{B(i \omega_k) \Sigma_n(i \omega_k)}{\left( \omega_k^2 + \tilde{m}^2 \right)^2 } \approx - \frac{1}{N} \frac{22}{9} \tilde{m}^2. \eeq
For convenience, we introduce the resummed chain of bubble diagrams $B(i \omega_n)$, which is calculated similarly to its real-time counterpart:
\beq \label{eq:bubble-chain-matsubara}
\begin{aligned}
B(i \omega_n) &= 1 -\frac{\lambda}{\beta} \sum_{\omega_k} \frac{1}{\omega_k^2 + \tilde{m}^2} \frac{1}{(\omega_n + \omega_k)^2 + \tilde{m}^2} B(i \omega_n) \\ &= 1 - \frac{\frac{\lambda}{\tilde{m}} \coth \frac{\beta \tilde{m}}{2}}{\omega_n^2 + 4 \tilde{m}^2 + \frac{\lambda}{\tilde{m}} \coth \frac{\beta \tilde{m}}{2}} \approx \frac{\omega_n^2 + 4 \tilde{m}^2}{\omega_n^2 + 6 \tilde{m}^2}.
\end{aligned} \eeq
Note that the approximate identities in Eqs.~\eqref{eq:self-energy-sun}--\eqref{eq:bubble-chain-matsubara} are valid only in the high-temperature limit, where $\tilde{m} \approx \sqrt[4]{\lambda/\beta}$. The self-energies~\eqref{eq:self-energy-sun}--\eqref{eq:self-energy-tadpole} are also straightforwardly calculated for arbitrary parameters of the model, but we do not reproduce here these general expressions due to their bulkiness.

Thus the $\mO(1/N)$ resummed Matsubara propagator in the high-temperature limit has the following form:
\beq \tilde{G}(i \omega_n) = \frac{1}{G(i \omega_n) - \Sigma(i \omega_n)} \approx \frac{1}{\omega_n^2 + \left( 1 - \frac{5}{9 N} \right) \tilde{m}^2 - \frac{1}{N} \frac{2}{3} \frac{\left( 7 \omega_n^2 + 25 \tilde{m}^2 \right) \tilde{m}^4}{\omega_n^4 + 14 \omega_n^2 \tilde{m}^2 + 25 \tilde{m}^4}}. \eeq
Analytically continuing this expression to real frequencies, we obtain the $\mO(1/N)$ resummed retarded propagator:
\beq i \tilde{G}^R(\omega) \approx \frac{i}{(\omega + i 0)^2 - \left( 1 - \frac{5}{9 N} \right) \tilde{m}^2 - \frac{\tilde{m}^4}{N} \frac{2}{3} \frac{7 (\omega + i 0)^2 - 25 \tilde{m}^2}{(\omega + i 0)^4 - 14 (\omega + i 0)^2 \tilde{m}^2 + 25 \tilde{m}^4}} \eeq
We emphasize that the retarded self-energy, $\Sigma^R(\omega) = \Sigma(i \omega_n \to \omega + i 0)$, does not contain any imaginary contributions. Hence, all six poles of the resummed retarded propagator are purely real:
\beq \begin{gathered}
\omega_{1,2} \approx \pm \, \tilde{m} \left( 1 -  \frac{7}{9} \frac{1}{N} \right), \\
\omega_{3,4} \approx \pm \, \tilde{m} \left( 1.4495 + \frac{0.2194}{N} \right), \\
\omega_{5,6} \approx \pm \, \tilde{m} \left( 3.4538 + \frac{0.0698}{N} \right),
\end{gathered} \eeq
where we neglect the $1/N^2$ terms that can be estimated only in higher orders of the $1/N$ expansion.

Therefore, the $1/N$ correction to the self-energy does not imply a conventional thermalization. In other words, the inverse dissipation time (which is proportional to the imaginary part of the retarded self-energy) in the quantum mechanical model~\eqref{eq:S} cannot be larger than $\Gamma \sim \tilde{m}/N^2$. In fact, we expect that for a finite $N$, this time is exactly zero because quantum mechanical systems with a finite number of degrees of freedom cannot thermalize at all\footnote{However, quantum mechanical systems with a large number of degrees of freedom can \textit{approximately} thermalize, i.e., mimic with a good accuracy the energy level distribution of a thermal system~\cite{DAlessio, Deutsch, Srednicki}. A famous example of such a model is the Sachdev-Ye-Kitaev model, which has a finite dissipation time $t_d = 1/\Gamma$ due to the chaotic nature of its interactions~\cite{Kitaev-talks, Maldacena-SYK, Kitaev, Sarosi, Rosenhaus, Trunin-SYK}.}. This is a crucial difference between the quantum mechanics and quantum field theory; e.g., compare the calculations of this appendix with the model~\cite{Swingle}.

\end{document}